%% file: main.tex
\documentclass{article}
\usepackage{neurosketch_preprint,times}
\usepackage[utf8]{inputenc}
\usepackage[T1]{fontenc}
\usepackage{amsmath,amsfonts}
\usepackage{booktabs,multirow,array,diagbox,makecell}
\usepackage{graphicx,nicefrac,microtype,xcolor}
\usepackage{enumitem}
\usepackage{flafter} % Keep floats after their source insertion point.
\usepackage{hyperref}
\usepackage{url}
\hypersetup{hidelinks}

\newcommand{\vpara}[1]{\noindent\textbf{#1 }}
\graphicspath{{figures/}}

\newsavebox{\NeuroSketchTableBox}
\newcommand{\NeuroSketchFitTable}[1]{%
  \sbox{\NeuroSketchTableBox}{#1}%
  \typeout{NEUROSKETCH TABLE \thetable: natural=\the\wd\NeuroSketchTableBox, limit=\the\linewidth}%
  \ifdim\wd\NeuroSketchTableBox>\linewidth
    \resizebox{\linewidth}{!}{\usebox{\NeuroSketchTableBox}}%
  \else
    \usebox{\NeuroSketchTableBox}%
  \fi
}

\title{NeuroSketch: A Practical Design Recipe for Neural Decoding}
\input{authors}

\makeatletter
\let\SpacingOriginalEndFloatBox\@endfloatbox
\def\@endfloatbox{\SpacingOriginalEndFloatBox\typeout{SPACING FLOAT \@captype: height=\the\ht\@currbox, depth=\the\dp\@currbox}}
\makeatother
\AtBeginDocument{\typeout{SPACING DEFAULTS: baseline=\the\baselineskip, parskip=\the\parskip, textfloatsep=\the\textfloatsep, floatsep=\the\floatsep, intextsep=\the\intextsep, captionabove=\the\abovecaptionskip, captionbelow=\the\belowcaptionskip, textheight=\the\textheight}}

\begin{document}
\maketitle
\begin{abstract}
  Neural decoding is fundamental to brain-computer interfaces, with growing applications in healthcare. Previous research has focused on leveraging signal processing and deep learning methods to enhance neural decoding performance. However, systematic guidance on architectural design for neural decoding remains limited. In this study, we develop NeuroSketch, a practical design recipe for neural decoding, through a basic architecture study followed by macro- and micro-level optimization. Comparing nine basic architectures, we find that CNN-2D outperforms other architectures in neural decoding tasks and explore its effectiveness from temporal and spatial perspectives. Building on this backbone, we combine gradual feature-map expansion and early downsampling at the macro level with grouped convolutions at the micro level. These choices form the recipe, which we instantiate as NeuroSketch-Base (1.4M parameters) and NeuroSketch-Large (4.2M parameters). The recipe is developed and evaluated through nearly 5,000 experiments across eight tasks spanning visual, auditory, and speech modalities and EEG, SEEG, and ECoG signals. Against ten baselines, the two variants collectively achieve the best accuracy on each task. Our code is available at \url{https://github.com/Galaxy-Dawn/NeuroSketch}.
\end{abstract}

\input{./text/01-introduction}

\input{./text/02-roadmap}
\input{./text/03-experiments}
\input{./text/04-conclusion_and_future_work}

% Declarations follow the main text.
\input{./text/00-statements}

\bibliographystyle{neurosketch_preprint}
\bibliography{references}

\clearpage
\appendix

\section*{Appendix Contents}
\begingroup
\small
\newcommand{\AppendixContentsEntry}[2]{%
  \par\noindent\hspace*{#1}%
  \parbox[t]{\dimexpr\linewidth-#1-2.2em\relax}{%
    \hyperref[#2]{\ref*{#2}\quad\nameref*{#2}}}%
  \hfill\hyperref[#2]{\pageref*{#2}}\par\vspace{3pt}%
}
\vspace{5pt}
\AppendixContentsEntry{0pt}{Related Work}
\vspace{5pt}
\AppendixContentsEntry{0pt}{Details of Datasets}
\AppendixContentsEntry{1.5em}{Dataset Description}
\AppendixContentsEntry{1.5em}{Data Preprocessing}
\vspace{5pt}
\AppendixContentsEntry{0pt}{Details of Models}
\AppendixContentsEntry{1.5em}{Models in the roadmap exploration}
\AppendixContentsEntry{1.5em}{NeuroSketch Implementation}
\AppendixContentsEntry{1.5em}{Baseline Models}
\vspace{5pt}
\AppendixContentsEntry{0pt}{Experimental Settings}
\AppendixContentsEntry{1.5em}{Data splitting}
\AppendixContentsEntry{1.5em}{Data augmentation}
\AppendixContentsEntry{1.5em}{Hyperparameters}
\AppendixContentsEntry{1.5em}{Compute resources}
\vspace{5pt}
\AppendixContentsEntry{0pt}{Experimental Analysis}
\AppendixContentsEntry{1.5em}{Scaling Behavior}
\AppendixContentsEntry{1.5em}{Representation Analysis of CNN-2D and Transformer‑Based Architectures}
\AppendixContentsEntry{1.5em}{Hyperparameter Analysis}
\AppendixContentsEntry{1.5em}{Component-wise Ablation Study}
\AppendixContentsEntry{1.5em}{Cross-Subject Evaluation}
\AppendixContentsEntry{1.5em}{Robustness to Reduced Training Data}
\AppendixContentsEntry{1.5em}{Computational Efficiency}
\endgroup
\clearpage
\section*{Appendix Overview}
The appendices provide supporting details for NeuroSketch, our practical design recipe, and supplementary evaluations of its two model instantiations. The guide below summarizes the purpose of each section.

\vpara{Appendix~\ref{Related Work}: Related work.}This appendix reviews neural decoding tasks and prior studies of neural architecture design.

\vpara{Appendix~\ref{Details of Datasets}: Details of Datasets.}Section~\ref{Dataset Description} summarizes the recording settings and experimental paradigms of the six datasets used in the main evaluation. Section~\ref{Data Preprocessing} describes how recordings are transformed into model inputs and classification targets.

\vpara{Appendix~\ref{Details of Models}: Details of Models.}Section~\ref{Models in the roadmap exploration} provides architectural details of the models examined in the roadmap study. Section~\ref{NeuroSketch Implementation} describes the shared architecture and configurations of NeuroSketch-Base and NeuroSketch-Large. Section~\ref{Baseline Models} summarizes the main architectural characteristics of the baseline models.

\vpara{Appendix~\ref{Experimental Settings}: Experimental settings.}Section~\ref{Data splitting} explains the training, validation, and test protocols for the roadmap study and main evaluation. Section~\ref{Data augmentation} describes the training-time transformations applied to neural signals. Section~\ref{Hyperparameters} documents the baseline configurations and total parameter counts, explains how the foundation models are adapted and fine-tuned, and presents the shared training hyperparameters and rationale for dataset-specific training durations. Section~\ref{Compute resources} reports the hardware, experiment counts, and average training times.

\vpara{Appendix~\ref{Experimental Analysis}: Additional experimental analyses.}Section~\ref{Scaling Behavior} examines how Base and Large performance differs across Du-IN subjects.
Section~\ref{Representation Analysis of CNN-2D and Transformer‑Based Architectures} explores differences in learned representations to better understand the performance gap between CNN-based and Transformer-based models. 
Section~\ref{Hyperparameter Analysis} assesses how sensitive NeuroSketch-Large is to convolutional kernel size and group count.
Section~\ref{Component-wise Ablation Study} examines how pooling choices and residual connections contribute to decoding performance.
Section~\ref{Cross-Subject Evaluation} evaluates NeuroSketch-Base's ability to generalize to participants unseen during training.
Section~\ref{Robustness to Reduced Training Data} examines whether NeuroSketch-Base retains its performance advantage when labeled training data are limited.
Section~\ref{Computational Efficiency} assesses the computational efficiency of both model instantiations during training and inference.

\clearpage

\input{./text/05-related_work}
\input{./text/06-appendix}
\end{document}

%% file: authors.tex
\author{%
  \begin{minipage}{\dimexpr\textwidth-2\tabcolsep\relax}
  \raggedright
  \textbf{Gaorui Zhang\textsuperscript{1}, Zhizhang Yuan\textsuperscript{1}, Jialan Yang\textsuperscript{1}, Junru Chen\textsuperscript{1}, Fanqi Shen\textsuperscript{1}, Li Meng\textsuperscript{2},\\ Yang Yang\textsuperscript{1,$\dagger$}}\\[4pt]
  \normalfont\small
  \textsuperscript{1}College of Computer Science and Technology, Zhejiang University\\
  \textsuperscript{2}Shanghai Institute of Microsystem and Information Technology\\[3pt]
  \texttt{\{gaoruizhang,zhizhangyuan,jlyang,jrchen\_cali,shenfanqi,yangya\}@zju.edu.cn}\\
  \texttt{li.meng@mail.sim.ac.cn}\\[3pt]
  \textsuperscript{$\dagger$}Corresponding author.
  \end{minipage}%
}
\hypersetup{pdfauthor={Gaorui Zhang, Zhizhang Yuan, Jialan Yang, Junru Chen, Fanqi Shen, Li Meng, Yang Yang}}

%% file: text/01-introduction.tex
\section{Introduction}
\label{introduction}
% Brain-Computer Interface (BCI) technology aims to revolutionize human-machine interaction by establishing direct pathways between neural activity and external action~\citep{Maiseli_2023_Brain}. 
% A critical component of BCI systems is neural decoding~\citep{VanGerven_2019_Current}, which infers external stimuli, cognitive states, or intentions from recorded brain signals.

Neural decoding infers stimuli, cognitive states, or intentions from recorded brain signals and is a central component of brain-computer interfaces~\citep{VanGerven_2019_Current}.
These signals are typically acquired using non-invasive methods such as electroencephalography (EEG)~\citep{Teplan_2002_Fundamentals} or invasive techniques such as stereoelectroencephalography (SEEG)~\citep{Talairach_1974_Approche} and electrocorticography (ECoG)~\citep{Shenoy_2007_Generalized}. While EEG offers practicality and low cost for research~\citep{Zhang_2021_survey, Altaheri_2023_Deep, Rahman_2021_Recognition} and clinical applications~\citep{Pani_2022_Clinical, Saminu_2023_Applications}, invasive methods provide superior temporal and spatial resolution by capturing signals from deeper brain structures~\citep{Chaddad_2023_Electroencephalography}.

In neural decoding, recorded brain signals are characterized by transient temporal dynamics~\citep{King_2014_Characterizing} and spatial locality~\citep{Mahjoory_2024_Convolutional}, which present unique modeling demands.
% In the data level, unlike traditional time series data such as weather, electricity consumption, and traffic flow, brain signals inherently exhibit high temporal resolution and non-stationarity.
Traditional time series data, such as weather, electricity consumption, and traffic flow, are sampled at relatively low frequencies (\emph{e.g.,} minutes or hours) and often exhibit significant periodicity or trends.
However, brain signals require higher temporal resolution to capture the brain's rapidly changing states.
Even within brain signals, their characteristics can differ significantly across various scenarios. From a temporal perspective, brain signals typically contain crucial information within much shorter time frames in neural decoding tasks compared to other scenarios, such as sleep staging. In sleep staging tasks, brain signals are recorded overnight, with a labeling resolution of 30 seconds for each sleep stage~\citep{Phan_2022_Automatica}. In contrast, in the Rapid Serial Visual Presentation (RSVP) paradigm for image stimulus decoding, relevant information is encoded within just a few hundred milliseconds~\citep{Butts_2007_Temporal}. 
% This temporal property necessitates models capable of identifying immediate and non-stationary neural responses to stimuli. 
From a spatial perspective, task-driven neural activity in decoding paradigms differs from activity associated with pathological conditions.
% Pathological conditions such as generalized epilepsy are characterized by widespread brain activity, where bilateral synchronous abnormal electrical discharges propagate across both hemispheres of the brain, resulting in extensive, non-specific channel interactions~\citep{Stafstrom_2015_Seizures}.
Pathological conditions such as generalized epilepsy exhibit widespread brain activity, resulting in non-specific neural activations~\citep{Stafstrom_2015_Seizures}. In neural decoding tasks, however, external stimuli evoke localized activations that are typically restricted to functionally specialized regions and recorded by spatially proximate electrodes. For example, speech stimuli predominantly activate the left inferior frontal gyrus~\citep{Leonard_2024_Large-scale}, while visual stimuli engage the visual cortex~\citep{Grill-Spector_2004_HUMAN}. 

Previous work on neural decoding has explored both signal processing and deep learning approaches. Signal processing methods aim to improve signal quality~\citep{Peksa_2023_State-of-the-art} and extract task-relevant features~\citep{Wittevrongel_2020_High‐gamma, Proix_2022_Imagined, Safi_2021_Early}, but manually designing these features requires substantial domain expertise and effort. Deep learning methods reduce this reliance on handcrafted features by learning task-relevant representations directly from neural signals~\citep{Angrick_2019_Speech, Makin_2020_Machine, Wilson_2020_Decoding, Willett_2023_high-performance, Metzger_2023_high-performance, Zheng_2024_Du-IN}.
% Previous work on neural decoding can be broadly categorized into two approaches. The first is signal processing~\citep{Peksa_2023_State-of-the-art}, which focuses on improving the signal-to-noise ratio and extracting task-relevant features~\citep{Wittevrongel_2020_High‐gamma, Proix_2022_Imagined, Safi_2021_Early}. These methods rely on manually defined features, which are time-consuming and require substantial domain expertise.
% The second approach involves deep learning methods~\citep{Angrick_2019_Speech, Makin_2020_Machine, Wilson_2020_Decoding, Willett_2023_high-performance, Metzger_2023_high-performance, Zheng_2024_Du-IN}.
% Due to the data scarcity of decoding tasks, these methods focus on end-to-end supervised learning paradigms for implicit feature extraction.
% However, the methods often involve directly deploying existing model architectures and methods, lacking in-depth exploration and modeling of the temporal dynamics and channel constraints inherent to neural decoding.
% However, existing approaches have largely overlooked in-depth architecture exploration. Therefore, our work aims to explore a framework more suited to neural decoding tasks through systematic architectural optimization.
However, systematic guidance on architectural design across diverse neural decoding tasks and recording types remains limited, despite the gains achieved through architectural studies in other domains~\citep{Yu_2022_Metaformer, Yu_2023_Metaformer, Luo_2024_Moderntcn, Wang_2025_TimeMixer}.
% However, an important aspect that remains largely overlooked is the in-depth exploration of model architecture, which also plays a crucial role in optimizing task performance.
For instance, in energy forecasting tasks, ModernTCN~\citep{Luo_2024_Moderntcn} reduces MSE by 13.9\% on ETTm2~\citep{Zhou_2021_Informer} compared to convolution-based models by using large kernel convolutions and a ConvFFN module. In image classification tasks, ConvNeXt~\citep{Liu_2022_convnet} surpasses ResNet-50~\citep{He_2016_Deep} on ImageNet~\citep{Deng_2009_Imagenet} through optimizations in stage compute ratio and convolutional methods. We therefore develop NeuroSketch, a practical design recipe for neural decoding.

% For example, in time series analysis, ModernTCN~\citep{Luo_2024_Moderntcn} achieves a 13.9\% reduction in MSE on ETTm2~\citep{Zhou_2021_Informer} compared to existing convolution-based models by adopting a large kernel convolution and a ConvFFN module. In computer vision, ConvNeXt~\citep{Liu_2022_convnet} outperforms traditional ResNet-50~\citep{He_2016_Deep} on ImageNet~\citep{Deng_2009_Imagenet} by optimizing the stage compute ratio, convolutional methods and other low-level components.
% Therefore, similar optimization for neural decoding tasks can bring substantial benefits.
To develop this recipe, we investigate two core questions:
% Our study is guided by two core questions that aim to systematically investigate how to design an effective framework for neural decoding:

\textit{Q1: Which basic model architecture is best suited to capture temporal and spatial patterns in brain signals?}

\textit{Q2: Based on an appropriate architecture, how can we improve neural decoding performance through macro-to-micro architectural optimization?}

% \begin{figure}[H]
%     \centering
%     \includegraphics[width=0.5\textwidth]{./figures/Fig1.pdf}
%     \caption{\textbf{Roadmap of architectural optimization.}}
%     \label{fig:Fig1}
% \end{figure}

\begin{figure}[htbp]
    \centering
    \includegraphics[width=\linewidth]{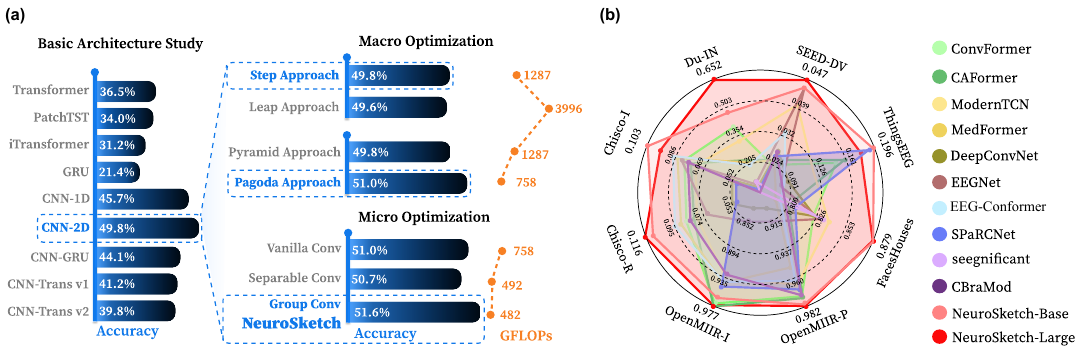}
    \caption{\textbf{Overview of NeuroSketch and its model instantiations.}
\textbf{(a)} Roadmap for developing the practical design recipe.
\textbf{(b)} Comparison of NeuroSketch-Base and NeuroSketch-Large with ten baselines across eight neural decoding tasks.}
    \label{fig:Fig1}
\end{figure}

As shown in Figure~\ref{fig:Fig1}(a), by answering the above questions, we consistently improve the decoding performance at each step. 
% First, we study 9 basic architectures (CNN~\citep{He_2016_Deep}, GRU~\citep{Chung_2014_Empirical}, Transformer~\citep{Vaswani_2017_Attention}, and their variants) and find that CNN-2D outperforms the others. The results are coherent with the characteristics of neural decoding signals, since CNN-2D is suitable for capturing localized patterns from both temporal and spatial dimensions. 
% Based on CNN-2D, we design the entire framework using a macro-to-micro paradigm, analyzing the latent space transformation during forward propagation and optimizing the calculation of micro components. 
We first study nine basic architectures (CNN~\citep{He_2016_Deep}, GRU~\citep{Chung_2014_Empirical}, Transformer~\citep{Vaswani_2017_Attention}, and their variants) and find that CNN-2D outperforms the others. This finding is consistent with the temporal and spatial locality of neural signals, which CNN-2D captures through local convolutional kernels.
Building on the backbone study, we further develop the recipe using a macro-to-micro paradigm, analyzing the latent space transformation during forward propagation and optimizing the calculation of micro components. Specifically, we find that gradual feature-map expansion and early downsampling at the macro level, together with grouped convolutions at the micro level, help balance decoding accuracy and computational cost. Taken together, the findings across all stages form the NeuroSketch recipe.
% Through the above step-by-step optimization process, we propose NeuroSketch, an effective framework for neural decoding. 
% NeuroSketch aligns with the temporal and spatial characteristics of brain signals in neural decoding tasks, achieving state-of-the-art performance across eight neural decoding tasks, which can serve as a useful tool for future research and applications in neuroscience.
We instantiate this recipe as NeuroSketch-Base and NeuroSketch-Large, with 1.4M and 4.2M parameters, respectively. The two variants collectively achieve the best or tied-best accuracy on all eight decoding tasks among the evaluated models (Figure~\ref{fig:Fig1}(b)), supporting the practical value of the recipe for neural decoding.

% Original standalone Figure 1 retained for reference.
% \begin{figure}[htbp]
%     \centering
%     \includegraphics[width=0.65\linewidth]{./figures/Fig1.pdf}
%     \caption{\textbf{Roadmap for developing a practical design recipe for neural decoding.}}
%     \label{fig:Fig1}
% \end{figure}

% \begin{wrapfigure}{r}{0.5\textwidth}
% \centering
% \includegraphics[width=0.5\textwidth]{./figures/Fig1.pdf}
% \caption{Roadmap of our optimization considering performance and computational cost.}
% \label{fig:Fig1}
% \end{wrapfigure}
% As shown in Figure~\ref{fig:Fig1}, by answering the above questions, we have achieved a consistent performance improvement at each optimization step. First, we study 9 basic architectures (CNN~\citep{He_2016_Deep}, GRU~\citep{Chung_2014_Empirical}, Transformer~\citep{Vaswani_2017_Attention}, and their variants) and find that CNN-2D outperforms the others, which is coherent with the characteristics of neural decoding signals. Additionally, we design the entire framework by a macro-to-micro paradigm, where we analyze the latent space transformation during forward propagation and optimize the calculation of micro components. Through the above step-by-step optimization process, we propose NeuroSketch, a efficient and effective framework for neural decoding. The design of NeuroSketch aligns with the temporal and spatial characteristics of neural decoding signals, achieving state-of-the-art performance and computational efficiency across eight mainstream neural decoding tasks. NeuroSketch can serve as a useful tool for future research and applications in neuroscience.

In summary, the main contributions of this research are as follows:
\begin{itemize}[leftmargin=*]
    \item [1.] We develop NeuroSketch, a practical design recipe for neural decoding, and instantiate it as NeuroSketch-Base and NeuroSketch-Large, tailored to the temporal and spatial characteristics of brain signals.
    \item [2.] From a technical perspective, we systematically reexamine the design space and consider several critical components. Starting with exploring basic architectures, we progressively investigate architectural design principles from macro- to micro-level.
    \item [3.] From an experimental perspective, we develop and evaluate the recipe through nearly 5,000 experiments across eight neural decoding tasks related to vision, hearing, and speech, recorded by different types of brain signals. The experimental results demonstrate that the two model instantiations collectively achieve state-of-the-art decoding performance across all evaluated datasets.
\end{itemize}

%% file: text/02-roadmap.tex
\section{Roadmap for Developing a Practical Design Recipe}\label{roadmap}
This section develops NeuroSketch, our practical design recipe for neural decoding. After introducing the experimental setup (Section~\ref{Experiment Setup}), we address Q1 through a basic architecture
  study (Section~\ref{Basic Architecture Study}). We then address Q2 through macro- and micro-level optimization (Sections~\ref{Latent Space Transformation} and~\ref{Optimization of Calculation}). Finally,
  Section~\ref{Put It All Together} consolidates these findings into the recipe and introduces its two instantiations, NeuroSketch-Base and NeuroSketch-Large.

\subsection{Experimental Setup}\label{Experiment Setup}
To develop the recipe, we conduct 1,764 experiments across eight neural decoding tasks spanning three modalities (visual, auditory, and speech) and three signal types (EEG, SEEG, and ECoG). Table~\ref{tab:Table.1} summarizes the datasets, with further details provided in Appendix~\ref{Details of Datasets}. To reduce differences in model size across architectures, we adjust network depth and embedding dimensions to keep parameter counts at approximately 30M. Detailed model configurations and training settings are provided in Appendices~\ref{Models in the roadmap exploration} and~\ref{Experimental Settings}.

\input{./tables/dataset_table}

\input{./tables/Q1_table_new}

\input{./tables/Q2_table}
\subsection{Basic Architecture Study}\label{Basic Architecture Study}

To address Q1, we compare nine architectures from four families: convolutional neural networks (CNNs), gated recurrent units (GRUs), Transformers, and hybrids. Each takes input $\mathbf{X}\in\mathbb{R}^{B\times C\times L}$, where $B$, $C$, and $L$ denote batch size, recording channels, and sequence length. CNN-1D combines all input channels through temporal filters, whereas CNN-2D reshapes the input to $\mathbf{X}^{\prime}\in\mathbb{R}^{B\times1\times C\times L}$ and applies local kernels along time and channel dimensions. The GRU sequentially updates its hidden state from the current input and previous state, while the Transformer models temporal dependencies through self-attention. We also evaluate PatchTST~\citep{Nie_2023_Time} and iTransformer~\citep{Liu_2024_iTransformer}. PatchTST divides each channel into patches of length $P$ and stride $S$, yielding $N=\left\lfloor(L-P)/S\right\rfloor+2$ patches. Transformer layers process the patched input $\mathbf{X}_p\in\mathbb{R}^{B\times C\times N\times P}$ independently across channels. In contrast, iTransformer treats each channel as a token to model inter-channel dependencies. Hybrids comprise CNN--GRU and two CNN--Transformer variants. The first CNN--Transformer stacks Transformer layers after a CNN feature extractor~\citep{Song_2022_EEG}, whereas the second combines convolution and self-attention within each block~\citep{Kim_2022_Squeezeformer}. CNN--GRU similarly stacks GRU layers after a CNN feature extractor.

Table~\ref{tab:Table.2} (upper) shows the performance of the 9 basic architectures discussed above. Overall, CNN-based models, which excel in capturing local patterns, outperform hybrid models on most neural decoding tasks. Meanwhile, GRU and Transformer-based models, which are better suited for modeling long-range dependencies, exhibit the lowest performance. This result aligns with the transient temporal dynamics of brain signals in neural decoding tasks, suggesting that short-range temporal information is more effective for neural decoding. 
% In CNN-based models, CNN-2D, which extracts local channel information using convolutional kernels, outperforms CNN-1D, which fuses all channel information through addition, which is consistent with the \textbf{spatially constrained channel interactions} in neural decoding tasks. 
Among CNN-based models, CNN-2D captures local cross-channel patterns, whereas CNN-1D aggregates information across all input channels. The higher accuracy of CNN-2D suggests that explicitly modeling local cross-channel dependencies is beneficial for neural decoding.
Surprisingly, the Transformer-based models perform poorly on certain datasets, particularly the ThingsEEG dataset, where their best accuracy of 0.6\% is only marginally above the chance level of 0.5\%. To understand the underlying reasons for these performance differences, we conduct further analyses from both temporal and spatial perspectives.

From a temporal perspective, we investigate how short- and long-term temporal information affects the model performance. We control the short- and long-term information through adjusting the ratio of CNN and Transformer layers in two hybrid models. From the results in Table~\ref{tab:Table.2} (middle), we observe a decline in performance as the proportion of short-term information decreases, demonstrating the effectiveness of short-range information for neural decoding.

\begin{figure}[tb]
   \centering
   \includegraphics[width=0.95\linewidth]{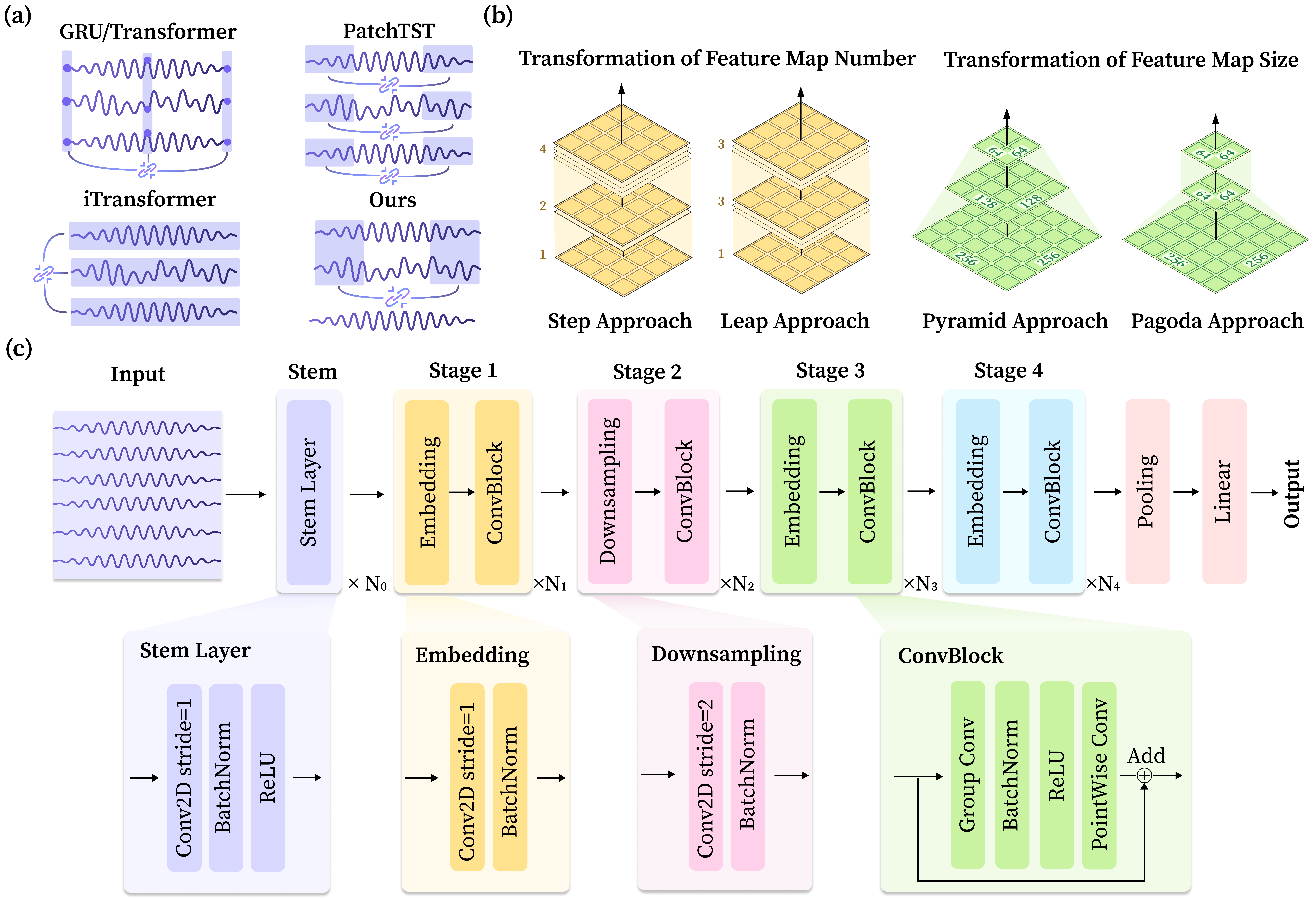}
   \caption{\textbf{(a) Comparison of patching schemes.} The vanilla GRU and Transformer use the values from all recording channels at one time step as a token; PatchTST uses a temporal window from a single channel; iTransformer uses an entire channel sequence. We also evaluate a patching scheme that jointly represents multiple time steps and recording channels. \textbf{(b) Comparison of different latent space transformation methods.} Regarding the number of feature maps, the step approach increases them gradually, whereas the leap approach increases them rapidly in the early stages. In terms of their size, the pyramid approach decreases them progressively, while the pagoda approach decreases them quickly in the early stages.
   \textbf{(c) The shared architecture of NeuroSketch-Base and NeuroSketch-Large.}} 
   % The input is first passed through a stem layer for initial processing, followed by four sequential stages for feature refinement. Each stage is composed of several layers, starting with an embedding or downsampling layer followed by a ConvBlock.
   \label{fig:Fig2}
\end{figure}

From a spatial perspective, we extend the CNN-2D finding to GRUs and Transformers through a patching scheme for local channel information. The input is divided into sliding temporal windows (Figure~\ref{fig:Fig2}(a)). Unlike PatchTST and iTransformer, which treat channels independently or focus on global channel information, the scheme concatenates channel and patch dimensions to preserve channel relationships within each window. This scheme improves Transformer accuracy on Du-IN from 8.5\% to 23.4\% (Table~\ref{tab:Table.2}, lower). GRU accuracy on OpenMIIR perception and imagination rises from chance level to 95.8\% and 93.7\%, respectively, further supporting spatial locality in neural decoding.

\textit{From now on, we will use CNN-2D as our basic architecture.}

\subsection{Macro-Level Design: Latent Space Transformation}\label{Latent Space Transformation}
To address the macro optimization aspect of Question 2, we study the latent space transformation during the forward propagation. For CNN-2D, its latent space representation refers to $\mathbf{S} \in \mathbb{R}^{C \times H \times W}$, where $C$ denotes the number of feature maps, $H$ and $W$ denote the height and width of feature maps, respectively. During forward propagation, the network increases the number of feature maps to capture higher-level information, while reducing the size of feature maps to alleviate redundancy and improve computational efficiency. Building on this general overview, we will investigate the transformation of the number ($C$) and the size ($H$ and $W$) of feature maps separately.

Regarding the number of feature maps, the appropriate strategy to increase them is a critical aspect of neural network design, and this can be achieved through two widely adopted approaches: \textbf{step approach} and \textbf{leap approach}. As shown in Figure{~\ref{fig:Fig2}}(b), the step approach gradually increases the number of feature maps, enabling the network to extract features from low to high levels systematically. In contrast, the leap approach rapidly increases the number of feature maps to the embedding dimension in the early stages, allowing the model to focus primarily on refining higher-level features.

The upper block of Table~\ref{tab:Table.3} compares the accuracy and computational cost of the step and leap approaches.
% Table{~\ref{tab:Table.3}}(upper) shows the performance and computational cost of the step approach and the leap approach. 
Surprisingly, the step approach achieves highly competitive performance compared to the leap approach, while requiring 67.8\% fewer FLOPs. 
On FacesHouses, the step approach achieves 88.6\% accuracy with 1,287 GFLOPs, compared with 85.7\% and 3,996 GFLOPs for the leap approach.
This indicates that neural decoding does not rely solely on high-level neural representations, and that multi-scale feature extraction is often more effective and efficient.

\textit{From now on, we will use the step approach to transform the number of feature maps.}

In the forward propagation process, the transformation of the feature map size is achieved through downsampling. Therefore, we further investigate how downsampling should be distributed throughout the network. Suppose downsampling is applied only in the final layers. In that case, earlier layers must process high-resolution feature maps, which significantly increases computational complexity and undermines the original purpose of downsampling to improve computational efficiency. Therefore, we study two common strategies: \textbf{pyramid approach} and \textbf{pagoda approach}. As shown in Figure{~\ref{fig:Fig2}}(b), the pyramid approach distributes the downsampling process evenly across the entire network, gradually reducing the size of the feature maps. This pyramid-like structure enables the network to retain more detailed features at the cost of increased computational complexity. The pagoda approach takes a more aggressive strategy by concentrating the downsampling module in the early stages of the network. This leads to a significant reduction in feature map size early on, resulting in lower computational costs. Table~\ref{tab:Table.3} (middle) compares the performance and computational cost between the pyramid and pagoda approach. The pagoda approach achieves slightly better accuracy than the pyramid approach, requiring 41.1\% fewer FLOPs. Specifically, on the Du-IN dataset, the pagoda approach attains 70.1\% accuracy, whereas the pyramid approach only reaches 64.7\% with a higher computational cost. On ThingsEEG, the pagoda approach increases accuracy by 2.0 percentage points to 20.4\%, exceeding CNN-1D's 20.2\%. These results demonstrate that excessive extraction of detailed information from the original signal is unnecessary. Brain signals typically have low signal-to-noise ratios, and early downsampling allows the model to focus on the most relevant features while reducing noise impact, thereby enhancing both computational efficiency and performance.

\textit{From now on, we will use the pagoda approach to transform the size of feature maps.} 

\subsection{Micro-Level Design: Convolutional Operations}\label{Optimization of Calculation}

To address the micro-level aspect of Q2, we compare convolutional operations. Let $\mathbf{X}_{\mathrm{in}}\in\mathbb{R}^{C_{\mathrm{in}}\times H_{\mathrm{in}}\times W_{\mathrm{in}}}$ be an input feature tensor. A standard convolution with $C_{\mathrm{out}}$ filters of size $U\times V$ produces an output tensor $\mathbf{X}_{\mathrm{out}}\in\mathbb{R}^{C_{\mathrm{out}}\times H_{\mathrm{out}}\times W_{\mathrm{out}}}$. Here, $C_{\mathrm{in}}$ and $C_{\mathrm{out}}$ denote the numbers of learned feature channels. The kernel tensor has shape $C_{\mathrm{out}}\times C_{\mathrm{in}}\times U\times V$, giving $C_{\mathrm{out}}C_{\mathrm{in}}UV$ kernel parameters when bias terms are excluded.

To explore more efficient alternatives to standard convolution, we consider grouped convolution~\citep{Ioannou_2017_Deep} and separable convolution~\citep{Howard_2017_Mobilenets}. 

\vpara{Grouped convolution.}With $G$ groups, each output channel connects to $C_{\mathrm{in}}/G$ input channels. The kernel tensor has shape $C_{\mathrm{out}}\times(C_{\mathrm{in}}/G)\times U\times V$, giving $C_{\mathrm{out}}C_{\mathrm{in}}UV/G$ kernel parameters, or $1/G$ of the standard convolution's kernel parameter count.

\vpara{Separable convolution.}Separable convolution combines depthwise spatial filtering with pointwise channel mixing. Depthwise convolution applies one filter per input channel ($G=C_{\mathrm{in}}$), with kernels $\mathbf{F}_d\in\mathbb{R}^{C_{\mathrm{in}}\times U\times V}$. Pointwise convolution mixes channels using $1\times1$ kernels $\mathbf{F}_p\in\mathbb{R}^{C_{\mathrm{out}}\times C_{\mathrm{in}}\times1\times1}$. Together, these operations use $C_{\mathrm{in}}UV+C_{\mathrm{in}}C_{\mathrm{out}}$ kernel parameters, corresponding to $\frac{1}{C_{\mathrm{out}}}+\frac{1}{UV}$ of the standard convolution's kernel parameter count.

The lower block of Table~\ref{tab:Table.3} compares the three convolutional variants. Grouped convolution achieves competitive accuracy at the lowest computational cost. On SEED-DV, standard convolution achieves 5.9\% accuracy at 758 GFLOPs, whereas separable convolution reaches 6.1\% with 35\% fewer FLOPs. At a similar cost, grouped convolution further improves accuracy to 6.9\%. These results support channel grouping as an effective way to aggregate local information while improving computational efficiency.

\textit{We will use group convolution as our core calculation method, completing the practical design recipe.}

\subsection{The Practical Design Recipe and Its Instantiations}\label{Put It All Together}

Synthesizing these findings, we obtain NeuroSketch, a practical design recipe for neural decoding. We instantiate it as NeuroSketch-Base and NeuroSketch-Large, with 1.4M and 4.2M parameters, respectively. Both share the architecture in Figure~\ref{fig:Fig2}(c), differing in stage widths and depths. A stem processes the reshaped 2D input, followed by four stages with gradual feature-map expansion and downsampling concentrated in the second stage. Each stage contains grouped convolutions, batch normalization, ReLU activations, and pointwise convolutions. GeM pooling~\citep{berman2019multigrain} aggregates the two feature-map dimensions, and a linear classifier produces the predictions. Appendix~\ref{NeuroSketch Implementation} provides implementation details.

%% file: tables/dataset_table.tex
\begingroup

\setlength{\tabcolsep}{4pt}   % Set the uniform column spacing
\renewcommand{\arraystretch}{1.0} % Reduce row height
\setlength{\heavyrulewidth}{1.0pt}   % Top and bottom line thickness
\setlength{\lightrulewidth}{0.5pt}    % Middle line thickness
\setlength{\arrayrulewidth}{0.5pt}    % Uniform thickness of horizontal lines

\begin{table}[htbp]
\centering
\caption{\textbf{Overview of the datasets used in our experiments.} We evaluate on six datasets in total. For \emph{Chisco} and \emph{OpenMIIR}, we partition them by task, yielding eight distinct tasks overall.}

\label{tab:Table.1}
\footnotesize
\NeuroSketchFitTable{\begin{tabular}{ccccccccc}
\toprule
Categories                & Datasets     & Tasks              & Signal Type 
\\ \midrule
\multirow{3}{*}{Speech}   & Du-IN~\citep{Zheng_2024_Du-IN}         & Loud Reading       & SEEG       \\
                          & Chisco-R~\citep{Zhang_2024_Chisco}     & Silent Reading     & EEG        \\
                          & Chisco-I~\citep{Zhang_2024_Chisco}     & Speech Imagination & EEG         \\
\midrule
\multirow{3}{*}{Visual}   & FacesHouses~\citep{Miller_2019_library}  & Binary Image Decoding     & ECoG        \\
                          & ThingsEEG~\citep{gifford2022large}    & Multi-Class Image Decoding     & EEG        \\
                          & SEED-DV~\citep{Liu_2024_EEG2video}      & Video Decoding     & EEG          \\
\midrule
\multirow{2}{*}{Auditory} & OpenMIIR-P~\citep{Stober_2015_music}   & Music Perception   & EEG        \\
                          & OpenMIIR-I~\citep{{Stober_2015_music}}   & Music Imagination  & EEG        \\
\bottomrule
\end{tabular}}
\end{table}
\endgroup

%% file: tables/Q1_table_new.tex
\begingroup
\setlength{\tabcolsep}{1.5pt}   % Set the uniform column spacing
\renewcommand{\arraystretch}{1.0} % Reduce row height

\setlength{\heavyrulewidth}{1.0pt}   % Top and bottom line thickness
\setlength{\lightrulewidth}{0.5pt}    % Middle line thickness
\setlength{\arrayrulewidth}{0.5pt}    % Uniform thickness of horizontal lines

\begin{table}[t]
\centering % 让表格居中
\caption{\textbf{Results of the basic architecture study and subsequent analyses.} v1 and v2 refer to the two hybrid methods discussed above. The ratio shown in the middle of the table (\emph{e.g.,} 4:1) indicates the proportion of CNN to Transformer layers in each hybrid model. The best results are in {\textbf{bold}} and the second best are {\underline{underlined}}.}
\label{tab:Table.2}
\NeuroSketchFitTable{ % 调整表格宽度为文本宽度，高度自适应
\fontsize{8}{9.5}\selectfont
\begin{tabular}{@{} p{0.9cm} p{0.55cm} p{0.55cm} p{0.55cm} p{0.55cm} p{0.55cm} p{0.55cm} p{0.55cm} p{0.55cm} p{0.55cm} p{0.55cm} p{0.55cm} p{0.55cm} p{0.55cm} p{0.52cm} p{0.52cm} p{0.52cm} p{0.52cm} @{}}
\toprule
\multicolumn{2}{c}{\multirow{2}{*}{\textbf{\diagbox{Models}{Datasets}}}} & \multicolumn{2}{c}{\textbf{Du-IN}} & \multicolumn{2}{c}{\textbf{SEED-DV}} & \multicolumn{2}{c}{\textbf{ThingsEEG}} & \multicolumn{2}{c}{\textbf{FacesHouses}} & \multicolumn{2}{c}{\textbf{OpenMIIR-P}} & \multicolumn{2}{c}{\textbf{OpenMIIR-I}} & \multicolumn{2}{c}{\textbf{Chisco-R}} & \multicolumn{2}{c}{\textbf{Chisco-I}}\\ 
\cmidrule(lr){3-4} \cmidrule(lr){5-6} \cmidrule(lr){7-8} \cmidrule(lr){9-10} \cmidrule(lr){11-12} \cmidrule(lr){13-14} \cmidrule(lr){15-16} \cmidrule(lr){17-18}
\multicolumn{2}{c}{} & \multicolumn{1}{c}{\textbf{Acc}} & \multicolumn{1}{c}{\textbf{F1}} & \multicolumn{1}{c}{\textbf{Acc}} & \multicolumn{1}{c}{\textbf{F1}} & \multicolumn{1}{c}{\textbf{Acc}} & \multicolumn{1}{c}{\textbf{F1}} & \multicolumn{1}{c}{\textbf{Acc}} & \multicolumn{1}{c}{\textbf{F1}} & \multicolumn{1}{c}{\textbf{Acc}} & \multicolumn{1}{c}{\textbf{F1}} & \multicolumn{1}{c}{\textbf{Acc}} & \multicolumn{1}{c}{\textbf{F1}} & \multicolumn{1}{c}{\textbf{Acc}} & \multicolumn{1}{c}{\textbf{F1}} & \multicolumn{1}{c}{\textbf{Acc}} & \multicolumn{1}{c}{\textbf{F1}}\\
\midrule
\multicolumn{18}{c}{\textbf{\textit{Basic Architecture Study}}} \\
\midrule
\multicolumn{2}{c}{CNN-1D}  
  & \multicolumn{1}{c}{{{\underline{.451}}}} % Col3: second highest (.451) vs .647 (max in CNN-2D)
  & \multicolumn{1}{c}{{{\underline{.446}}}} % Col4: second highest (.446) vs .641 (max in CNN-2D)
  & \multicolumn{1}{c}{.061}            % Col5: .061 vs .063 (underline in CNN-2D) vs .067 (max in CNN-Trans v1)
  & \multicolumn{1}{c}{.054}            % Col6: .054 vs .058 (underline in CNN-2D) vs .060 (max in CNN-Trans v1)
  & \multicolumn{1}{c}{{{\textbf{.202}}}}   % Col7: max (.202) from CNN-1D
  & \multicolumn{1}{c}{{{\textbf{.191}}}}   % Col8: max (.191) from CNN-1D
  & \multicolumn{1}{c}{.799}            % Col9: .799 vs {\underline{.845} (PatchTST) vs .886 (max in CNN-2D)
  & \multicolumn{1}{c}{.796}            % Col10: .796 vs {\underline{.838} (PatchTST) vs .885 (max in CNN-2D)
  & \multicolumn{1}{c}{{{\underline{.973}}}} % Col11: second highest (.973 from CNN-1D) vs .981 (max in CNN-2D)
  & \multicolumn{1}{c}{{{\underline{.973}}}} % Col12: second highest (.973 from CNN-1D) vs .980 (max in CNN-2D)
  & \multicolumn{1}{c}{{\underline{.968}}} % Col13: second highest (.968 from CNN-1D) vs .982 (max in CNN-2D)
  & \multicolumn{1}{c}{{\underline{.968}}} % Col14: second highest (.968 from CNN-1D) vs .982 (max in CNN-2D)
  & \multicolumn{1}{c}{.104}            % Col15: .104 vs .127 (max in CNN-2D) vs .123 (GRU gets underline)
  & \multicolumn{1}{c}{.071}            % Col16: .071 vs .113 (max in CNN-2D) vs .091 (GRU gets underline)
  & \multicolumn{1}{c}{.100}            % Col17: .100 vs {\underline{.114} (CNN-2D) vs .117 (max in GRU)
  & \multicolumn{1}{c}{.073}            % Col18: .073 vs .095 (max in CNN-2D) vs .086 (GRU gets underline)
  \\

\multicolumn{2}{c}{\textbf{CNN-2D}}
  & \multicolumn{1}{c}{{\textbf{.647}}}   % Col3: max (.647)
  & \multicolumn{1}{c}{{\textbf{.641}}}   % Col4: max (.641)
  & \multicolumn{1}{c}{{\underline{.063}}} % Col5: second highest (.063 from CNN-2D) vs .067 (max in CNN-Trans v1)
  & \multicolumn{1}{c}{{\underline{.058}}} % Col6: second highest (.058 from CNN-2D) vs .060 (max in CNN-Trans v1)
  & \multicolumn{1}{c}{{\underline{.184}}}            % Col7: .184 becomes underline later if needed; see below.
                                         % (For Col7, max is .202 in CNN-1D so here .184 is second highest.)
                                         % Thus, we underline it:
  & \multicolumn{1}{c}{{\underline{.177}}} % Col8: second highest (.177 from CNN-2D) vs .191 (max in CNN-1D)
  & \multicolumn{1}{c}{{\textbf{.886}}}   % Col9: max (.886)
  & \multicolumn{1}{c}{{\textbf{.885}}}   % Col10: max (.885)
  & \multicolumn{1}{c}{{\textbf{.981}}}   % Col11: max (.981)
  & \multicolumn{1}{c}{{\textbf{.980}}}   % Col12: max (.980)
  & \multicolumn{1}{c}{{\textbf{.982}}}   % Col13: max (.982)
  & \multicolumn{1}{c}{{\textbf{.982}}}   % Col14: max (.982)
  & \multicolumn{1}{c}{{\textbf{.127}}}   % Col15: max (.127)
  & \multicolumn{1}{c}{{\textbf{.113}}}   % Col16: max (.113)
  & \multicolumn{1}{c}{{\underline{.114}}} % Col17: second highest (.114 from CNN-2D) vs .117 (max in GRU)
  & \multicolumn{1}{c}{{\textbf{.095}}}   % Col18: max (.095)
  \\

\multicolumn{2}{c}{GRU}     
  & \multicolumn{1}{c}{.353}            
  & \multicolumn{1}{c}{.343}            
  & \multicolumn{1}{c}{.025}            
  & \multicolumn{1}{c}{.001}            
  & \multicolumn{1}{c}{.040}            
  & \multicolumn{1}{c}{.037}            
  & \multicolumn{1}{c}{.774}            
  & \multicolumn{1}{c}{.773}            
  & \multicolumn{1}{c}{.146}            
  & \multicolumn{1}{c}{.046}            
  & \multicolumn{1}{c}{.136}            
  & \multicolumn{1}{c}{.039}            
  & \multicolumn{1}{c}{{\underline{.123}}} % Col15: second highest (.123 from GRU) vs .127 (max in CNN-2D)
  & \multicolumn{1}{c}{{\underline{.091}}} % Col16: second highest (.091 from GRU) vs .113 (max in CNN-2D)
  & \multicolumn{1}{c}{{\textbf{.117}}}   % Col17: max (.117)
  & \multicolumn{1}{c}{{\underline{.086}}} % Col18: second highest (.086 from GRU) vs .095 (max in CNN-2D)
  \\

\multicolumn{2}{c}{Transformer}  
  & \multicolumn{1}{c}{.085}            
  & \multicolumn{1}{c}{.080}            
  & \multicolumn{1}{c}{.042}            
  & \multicolumn{1}{c}{.028}            
  & \multicolumn{1}{c}{.005}            
  & \multicolumn{1}{c}{.000}            
  & \multicolumn{1}{c}{.795}            
  & \multicolumn{1}{c}{.794}            
  & \multicolumn{1}{c}{.953}            
  & \multicolumn{1}{c}{.952}            
  & \multicolumn{1}{c}{.911}            
  & \multicolumn{1}{c}{.910}            
  & \multicolumn{1}{c}{.069}            
  & \multicolumn{1}{c}{.015}            
  & \multicolumn{1}{c}{.057}            
  & \multicolumn{1}{c}{.012}            
  \\

\multicolumn{2}{c}{PatchTST}  
  & \multicolumn{1}{c}{.060}            
  & \multicolumn{1}{c}{.051}            
  & \multicolumn{1}{c}{.038}            
  & \multicolumn{1}{c}{.029}            
  & \multicolumn{1}{c}{.006}            
  & \multicolumn{1}{c}{.001}            
  & \multicolumn{1}{c}{{\underline{.845}}} % Col9: second highest (.845 from PatchTST) vs .886 (max in CNN-2D)
  & \multicolumn{1}{c}{{\underline{.838}}} % Col10: second highest (.838 from PatchTST) vs .885 (max in CNN-2D)
  & \multicolumn{1}{c}{.906}            
  & \multicolumn{1}{c}{.904}            
  & \multicolumn{1}{c}{.766}            
  & \multicolumn{1}{c}{.760}            
  & \multicolumn{1}{c}{.050}             
  & \multicolumn{1}{c}{.005}            
  & \multicolumn{1}{c}{.050}            
  & \multicolumn{1}{c}{.005}            
  \\

\multicolumn{2}{c}{iTransformer}  
  & \multicolumn{1}{c}{.097}            
  & \multicolumn{1}{c}{.085}            
  & \multicolumn{1}{c}{.029}            
  & \multicolumn{1}{c}{.005}            
  & \multicolumn{1}{c}{.005}            
  & \multicolumn{1}{c}{.000}            
  & \multicolumn{1}{c}{.730}            
  & \multicolumn{1}{c}{.728}            
  & \multicolumn{1}{c}{.729}            
  & \multicolumn{1}{c}{.698}            
  & \multicolumn{1}{c}{.780}            
  & \multicolumn{1}{c}{.774}            
  & \multicolumn{1}{c}{.064}            
  & \multicolumn{1}{c}{.013}            
  & \multicolumn{1}{c}{.061}            
  & \multicolumn{1}{c}{.014}            
  \\

\multicolumn{2}{c}{CNN-GRU}  
  & \multicolumn{1}{c}{.416}            
  & \multicolumn{1}{c}{.409}            
  & \multicolumn{1}{c}{.057}            
  & \multicolumn{1}{c}{.049}            
  & \multicolumn{1}{c}{.172}            
  & \multicolumn{1}{c}{.166}            
  & \multicolumn{1}{c}{.803}            
  & \multicolumn{1}{c}{.801}            
  & \multicolumn{1}{c}{.957}            
  & \multicolumn{1}{c}{.954}            
  & \multicolumn{1}{c}{.943}            
  & \multicolumn{1}{c}{.942}            
  & \multicolumn{1}{c}{.092}            
  & \multicolumn{1}{c}{.064}            
  & \multicolumn{1}{c}{.087}            
  & \multicolumn{1}{c}{.063}            
  \\

\multicolumn{2}{c}{CNN-Trans v1}  
  & \multicolumn{1}{c}{.333}            
  & \multicolumn{1}{c}{.326}            
  & \multicolumn{1}{c}{{\textbf{.067}}}   % Col5: max (.067 from CNN-Trans v1)
  & \multicolumn{1}{c}{{\textbf{.060}}}   % Col6: max (.060 from CNN-Trans v1)
  & \multicolumn{1}{c}{.046}            
  & \multicolumn{1}{c}{.027}            
  & \multicolumn{1}{c}{.752}            
  & \multicolumn{1}{c}{.751}            
  & \multicolumn{1}{c}{.948}            
  & \multicolumn{1}{c}{.947}            
  & \multicolumn{1}{c}{.945}            
  & \multicolumn{1}{c}{.945}            
  & \multicolumn{1}{c}{.106}            
  & \multicolumn{1}{c}{.061}            
  & \multicolumn{1}{c}{.100}            
  & \multicolumn{1}{c}{.057}            
  \\

\multicolumn{2}{c}{CNN-Trans v2}  
  & \multicolumn{1}{c}{.282}            
  & \multicolumn{1}{c}{.268}            
  & \multicolumn{1}{c}{.050}            
  & \multicolumn{1}{c}{.045}            
  & \multicolumn{1}{c}{.022}            
  & \multicolumn{1}{c}{.010}            
  & \multicolumn{1}{c}{.754}            
  & \multicolumn{1}{c}{.752}            
  & \multicolumn{1}{c}{.952}            
  & \multicolumn{1}{c}{.950}            
  & \multicolumn{1}{c}{.930}            
  & \multicolumn{1}{c}{.930}            
  & \multicolumn{1}{c}{.097}            
  & \multicolumn{1}{c}{.050}            
  & \multicolumn{1}{c}{.093}            
  & \multicolumn{1}{c}{.059}            
  \\
\midrule

\multicolumn{18}{c}{\textbf{\textit{Ratio of CNN and Transformer Layers in Hybrid Architectures}}} \\ 
\midrule  
\multicolumn{2}{c}{\textbf{CNN-Trans v1-4:1}} 
  & \multicolumn{1}{c}{{\textbf{.333}}} 
  & \multicolumn{1}{c}{{\textbf{.326}}} 
  & \multicolumn{1}{c}{{\underline{.067}}}
  & \multicolumn{1}{c}{{\textbf{.060}}} 
  & \multicolumn{1}{c}{{\textbf{.046}}} 
  & \multicolumn{1}{c}{{\textbf{.027}} }
  & \multicolumn{1}{c}{{\textbf{.752}} }
  & \multicolumn{1}{c}{{\textbf{.751}} }
  & \multicolumn{1}{c}{{\textbf{.948}} }
  & \multicolumn{1}{c}{{\textbf{.947}} }
  & \multicolumn{1}{c}{{\textbf{.945}} }
  & \multicolumn{1}{c}{{\textbf{.945}} }
  & \multicolumn{1}{c}{{\textbf{.106}} }
  & \multicolumn{1}{c}{{\textbf{.061}} }
  & \multicolumn{1}{c}{{\textbf{.100}} }
  & \multicolumn{1}{c}{{\textbf{.057}}} \\ 
\multicolumn{2}{c}{CNN-Trans v1-3:2} 
  & \multicolumn{1}{c}{{\underline{.115}}} 
  & \multicolumn{1}{c}{{\underline{.091}}} 
  & \multicolumn{1}{c}{{\textbf{.069}}} 
  & \multicolumn{1}{c}{{\underline{.057}}} 
  & \multicolumn{1}{c}{{\underline{.010}}} 
  & \multicolumn{1}{c}{{\underline{.001}}} 
  & \multicolumn{1}{c}{.696} 
  & \multicolumn{1}{c}{.694} 
  & \multicolumn{1}{c}{{\underline{.904}}} 
  & \multicolumn{1}{c}{{\underline{.901}}} 
  & \multicolumn{1}{c}{{\underline{.916}}}
  & \multicolumn{1}{c}{{\underline{.914}}}
  & \multicolumn{1}{c}{{\underline{.094}}}
  & \multicolumn{1}{c}{{\underline{.036}}}
  & \multicolumn{1}{c}{{\underline{.090}}}
  & \multicolumn{1}{c}{{\underline{.041}}} \\ 
\multicolumn{2}{c}{CNN-Trans v1-2:3} 
  & \multicolumn{1}{c}{.060} 
  & \multicolumn{1}{c}{.032} 
  & \multicolumn{1}{c}{.041} 
  & \multicolumn{1}{c}{.021} 
  & \multicolumn{1}{c}{.005} 
  & \multicolumn{1}{c}{.000} 
  & \multicolumn{1}{c}{{\underline{.699}}} 
  & \multicolumn{1}{c}{{\underline{.697}}} 
  & \multicolumn{1}{c}{.833} 
  & \multicolumn{1}{c}{.828} 
  & \multicolumn{1}{c}{.867} 
  & \multicolumn{1}{c}{.866} 
  & \multicolumn{1}{c}{.088} 
  & \multicolumn{1}{c}{.030} 
  & \multicolumn{1}{c}{.085} 
  & \multicolumn{1}{c}{.031} \\ 
\multicolumn{2}{c}{CNN-Trans v1-1:4} 
  & \multicolumn{1}{c}{.036} 
  & \multicolumn{1}{c}{.013} 
  & \multicolumn{1}{c}{.030} 
  & \multicolumn{1}{c}{.008} 
  & \multicolumn{1}{c}{.005} 
  & \multicolumn{1}{c}{.000} 
  & \multicolumn{1}{c}{.684} 
  & \multicolumn{1}{c}{.674} 
  & \multicolumn{1}{c}{.709} 
  & \multicolumn{1}{c}{.687} 
  & \multicolumn{1}{c}{.764} 
  & \multicolumn{1}{c}{.760} 
  & \multicolumn{1}{c}{.079} 
  & \multicolumn{1}{c}{.018} 
  & \multicolumn{1}{c}{.076} 
  & \multicolumn{1}{c}{.019} \\ 
\midrule

\multicolumn{2}{c}{\textbf{CNN-Trans v2-4:1}} 
  & \multicolumn{1}{c}{{\textbf{.282}}} 
  & \multicolumn{1}{c}{{\textbf{.268}}} 
  & \multicolumn{1}{c}{.050} 
  & \multicolumn{1}{c}{.045} 
  & \multicolumn{1}{c}{{\underline{.022}}} 
  & \multicolumn{1}{c}{{\textbf{.010}}} 
  & \multicolumn{1}{c}{{\textbf{.754}} }
  & \multicolumn{1}{c}{{\textbf{.752}} }
  & \multicolumn{1}{c}{{\textbf{.952}} }
  & \multicolumn{1}{c}{{\textbf{.950}} }
  & \multicolumn{1}{c}{{\textbf{.930}} }
  & \multicolumn{1}{c}{{\textbf{.930}} }
  & \multicolumn{1}{c}{{\textbf{.097}} }
  & \multicolumn{1}{c}{.{\underline{050}}} 
  & \multicolumn{1}{c}{{\textbf{.093}}} 
  & \multicolumn{1}{c}{{\textbf{.059}}} \\ 
\multicolumn{2}{c}{CNN-Trans v2-3:2} 
  & \multicolumn{1}{c}{{\underline{.211}}} 
  & \multicolumn{1}{c}{{\underline{.192}}} 
  & \multicolumn{1}{c}{{\textbf{.061}}} 
  & \multicolumn{1}{c}{{\textbf{.054}}} 
  & \multicolumn{1}{c}{{\textbf{.024}}} 
  & \multicolumn{1}{c}{{\underline{.009}}} 
  & \multicolumn{1}{c}{.719} 
  & \multicolumn{1}{c}{.713} 
  & \multicolumn{1}{c}{{\underline{.946}}} 
  & \multicolumn{1}{c}{{\underline{.944}}} 
  & \multicolumn{1}{c}{.882} 
  & \multicolumn{1}{c}{.880} 
  & \multicolumn{1}{c}{.097}
  & \multicolumn{1}{c}{{\textbf{.057}}} 
  & \multicolumn{1}{c}{.087} 
  & \multicolumn{1}{c}{{\underline{.051}}} \\ 
\multicolumn{2}{c}{CNN-Trans v2-2:3} 
  & \multicolumn{1}{c}{.141} 
  & \multicolumn{1}{c}{.124} 
  & \multicolumn{1}{c}{{\underline{.059}}} 
  & \multicolumn{1}{c}{{\underline{.046}}} 
  & \multicolumn{1}{c}{.011} 
  & \multicolumn{1}{c}{.002} 
  & \multicolumn{1}{c}{.710} 
  & \multicolumn{1}{c}{.707} 
  & \multicolumn{1}{c}{.882} 
  & \multicolumn{1}{c}{.881} 
  & \multicolumn{1}{c}{{\underline{.893}}} 
  & \multicolumn{1}{c}{{\underline{.892}}} 
  & \multicolumn{1}{c}{.088} 
  & \multicolumn{1}{c}{.050} 
  & \multicolumn{1}{c}{{\underline{.092}}} 
  & \multicolumn{1}{c}{.044} \\ 
\multicolumn{2}{c}{CNN-Trans v2-1:4} 
  & \multicolumn{1}{c}{.047} 
  & \multicolumn{1}{c}{.033} 
  & \multicolumn{1}{c}{.035} 
  & \multicolumn{1}{c}{.019} 
  & \multicolumn{1}{c}{.008} 
  & \multicolumn{1}{c}{.001} 
  & \multicolumn{1}{c}{{\underline{.722}}} 
  & \multicolumn{1}{c}{{\underline{.720}}} 
  & \multicolumn{1}{c}{.822} 
  & \multicolumn{1}{c}{.814} 
  & \multicolumn{1}{c}{.839} 
  & \multicolumn{1}{c}{.837} 
  & \multicolumn{1}{c}{{\underline{.094}}} 
  & \multicolumn{1}{c}{.047} 
  & \multicolumn{1}{c}{.084} 
  & \multicolumn{1}{c}{.041} \\ 
\midrule

\multicolumn{18}{c}{\textbf{\textit{New Patch Method for Transformers and GRUs}}} \\ 
\midrule
\multicolumn{2}{c}{Transformer}      
  & \multicolumn{1}{c}{.085} 
  & \multicolumn{1}{c}{.080} 
  & \multicolumn{1}{c}{{\textbf{.042}}} 
  & \multicolumn{1}{c}{{\underline{.028}}} 
  & \multicolumn{1}{c}{{\underline{.005}}} 
  & \multicolumn{1}{c}{{\underline{.000}}} 
  & \multicolumn{1}{c}{.795} 
  & \multicolumn{1}{c}{.794} 
  & \multicolumn{1}{c}{{\textbf{.953}}} 
  & \multicolumn{1}{c}{{\textbf{.952}}} 
  & \multicolumn{1}{c}{{\textbf{.911}}} 
  & \multicolumn{1}{c}{{\textbf{.910}}} 
  & \multicolumn{1}{c}{{\underline{.069}}} 
  & \multicolumn{1}{c}{{\underline{.015}}} 
  & \multicolumn{1}{c}{.057} 
  & \multicolumn{1}{c}{.012} \\ 

\multicolumn{2}{c}{PatchTST}     
  & \multicolumn{1}{c}{.060} 
  & \multicolumn{1}{c}{.051} 
  & \multicolumn{1}{c}{{\underline{.038}}} 
  & \multicolumn{1}{c}{{\textbf{.029}}} 
  & \multicolumn{1}{c}{{\textbf{.006}}} 
  & \multicolumn{1}{c}{{\textbf{.001}}} 
  & \multicolumn{1}{c}{{\textbf{.845}}} 
  & \multicolumn{1}{c}{{\textbf{.838}}} 
  & \multicolumn{1}{c}{.906} 
  & \multicolumn{1}{c}{.904} 
  & \multicolumn{1}{c}{.766} 
  & \multicolumn{1}{c}{.760} 
  & \multicolumn{1}{c}{.050} 
  & \multicolumn{1}{c}{.005} 
  & \multicolumn{1}{c}{.050} 
  & \multicolumn{1}{c}{.005} \\ 

\multicolumn{2}{c}{iTransformer} 
  & \multicolumn{1}{c}{{\underline{.097}}} 
  & \multicolumn{1}{c}{{\underline{.085}}} 
  & \multicolumn{1}{c}{.029} 
  & \multicolumn{1}{c}{.005} 
  & \multicolumn{1}{c}{.005} 
  & \multicolumn{1}{c}{.000} 
  & \multicolumn{1}{c}{.730} 
  & \multicolumn{1}{c}{.728} 
  & \multicolumn{1}{c}{.729} 
  & \multicolumn{1}{c}{.698} 
  & \multicolumn{1}{c}{.780} 
  & \multicolumn{1}{c}{.774} 
  & \multicolumn{1}{c}{.064} 
  & \multicolumn{1}{c}{.013} 
  & \multicolumn{1}{c}{{\underline{.061}}} 
  & \multicolumn{1}{c}{{\underline{.014}}} \\ 

\multicolumn{2}{c}{\textbf{Ours}}         
  & \multicolumn{1}{c}{{\textbf{.234}}} 
  & \multicolumn{1}{c}{{\textbf{.226}}} 
  & \multicolumn{1}{c}{.033} 
  & \multicolumn{1}{c}{.027} 
  & \multicolumn{1}{c}{.005} 
  & \multicolumn{1}{c}{.000} 
  & \multicolumn{1}{c}{{\underline{.799}}} 
  & \multicolumn{1}{c}{{\underline{.799}}}
  & \multicolumn{1}{c}{{\underline{.922}}} 
  & \multicolumn{1}{c}{{\underline{.919}}} 
  & \multicolumn{1}{c}{{\underline{.843}}} 
  & \multicolumn{1}{c}{{\underline{.839}}} 
  & \multicolumn{1}{c}{{\textbf{.098}}} 
  & \multicolumn{1}{c}{{\textbf{.046}}} 
  & \multicolumn{1}{c}{{\textbf{.090}}} 
  & \multicolumn{1}{c}{{\textbf{.066}}} \\ 
\midrule

\multicolumn{2}{c}{GRU}      
  & \multicolumn{1}{c}{{\textbf{.353}}} % Col1: max = .353 vs .351
  & \multicolumn{1}{c}{{\textbf{.343}}} % Col2: max = .343 vs .342
  & \multicolumn{1}{c}{.025}         % Col3: max = .033 (Ours)
  & \multicolumn{1}{c}{.001}         % Col4: max = .026 (Ours)
  & \multicolumn{1}{c}{.040}         % Col5: max = .074 (Ours)
  & \multicolumn{1}{c}{.037}         % Col6: max = .068 (Ours)
  & \multicolumn{1}{c}{.774}         % Col7: max = .817 (Ours)
  & \multicolumn{1}{c}{.773}         % Col8: max = .813 (Ours)
  & \multicolumn{1}{c}{.146}         % Col9: max = .958 (Ours)
  & \multicolumn{1}{c}{.046}         % Col10: max = .956 (Ours)
  & \multicolumn{1}{c}{.136}         % Col11: max = .937 (Ours)
  & \multicolumn{1}{c}{.039}         % Col12: max = .936 (Ours)
  & \multicolumn{1}{c}{{\textbf{.123}}} % Col13: max = .123 vs .105
  & \multicolumn{1}{c}{{\textbf{.091}}} % Col14: max = .091 vs .071
  & \multicolumn{1}{c}{{\textbf{.117}}} % Col15: max = .117 vs .091
  & \multicolumn{1}{c}{{\textbf{.086}}} % Col16: max = .086 vs .066
  \\
\multicolumn{2}{c}{\textbf{Ours}} 
  & \multicolumn{1}{c}{.351}         
  & \multicolumn{1}{c}{.342}         
  & \multicolumn{1}{c}{{\textbf{.033}}} % Col3: max = .033
  & \multicolumn{1}{c}{{\textbf{.026}}} % Col4: max = .026
  & \multicolumn{1}{c}{{\textbf{.074}}} % Col5: max = .074
  & \multicolumn{1}{c}{{\textbf{.068}}} % Col6: max = .068
  & \multicolumn{1}{c}{{\textbf{.817}}} % Col7: max = .817
  & \multicolumn{1}{c}{{\textbf{.813}}} % Col8: max = .813
  & \multicolumn{1}{c}{{\textbf{.958}}} % Col9: max = .958
  & \multicolumn{1}{c}{{\textbf{.956}}} % Col10: max = .956
  & \multicolumn{1}{c}{{\textbf{.937}}} % Col11: max = .937
  & \multicolumn{1}{c}{{\textbf{.936}}} % Col12: max = .936
  & \multicolumn{1}{c}{.105}         
  & \multicolumn{1}{c}{.071}         
  & \multicolumn{1}{c}{.091}         
  & \multicolumn{1}{c}{.066}         
  \\ 
\bottomrule
\end{tabular}}
\end{table}

\endgroup

%% file: tables/Q2_table.tex
\begingroup
\setlength{\tabcolsep}{1.0pt}   % Set the uniform column spacing
\renewcommand{\arraystretch}{1.0} % Reduce row height

\setlength{\heavyrulewidth}{1.0pt}   % Top and bottom line thickness
\setlength{\lightrulewidth}{0.5pt}    % Middle line thickness
\setlength{\arrayrulewidth}{0.5pt}    % Uniform thickness of horizontal lines

\begin{table}[htbp]
\centering % 让表格居中
\caption{\textbf{Macro- and micro-level studies for developing the practical design recipe.} Lower GFLOPs indicates lower computational cost. The best results are in {{\textbf{bold}}}.}
\label{tab:Table.3}
\NeuroSketchFitTable{ % 调整表格宽度为文本宽度，高度自适应
\fontsize{8}{9.5}\selectfont
\begin{tabular}{@{}lc*{16}{c}@{}}
\toprule
\multirow{2}{*}{\textbf{Setting}} & \multicolumn{1}{c}{\multirow{2}{*}{\textbf{\diagbox{GFLOPs}{Datasets}}}} & \multicolumn{2}{c}{\textbf{Du-IN}} & \multicolumn{2}{c}{\textbf{SEED-DV}} & \multicolumn{2}{c}{\textbf{ThingsEEG}} & \multicolumn{2}{c}{\textbf{FacesHouses}} & \multicolumn{2}{c}{\textbf{OpenMIIR-P}} & \multicolumn{2}{c}{\textbf{OpenMIIR-I}} & \multicolumn{2}{c}{\textbf{Chisco-R}} & \multicolumn{2}{c}{\textbf{Chisco-I}}\\ 
\cmidrule(lr){3-4} \cmidrule(lr){5-6} \cmidrule(lr){7-8} \cmidrule(lr){9-10} \cmidrule(lr){11-12} \cmidrule(lr){13-14} \cmidrule(lr){15-16} \cmidrule(lr){17-18}
\multicolumn{2}{c}{} & \multicolumn{1}{c}{\textbf{Acc}} & \multicolumn{1}{c}{\textbf{F1}} & \multicolumn{1}{c}{\textbf{Acc}} & \multicolumn{1}{c}{\textbf{F1}} & \multicolumn{1}{c}{\textbf{Acc}} & \multicolumn{1}{c}{\textbf{F1}} & \multicolumn{1}{c}{\textbf{Acc}} & \multicolumn{1}{c}{\textbf{F1}} & \multicolumn{1}{c}{\textbf{Acc}} & \multicolumn{1}{c}{\textbf{F1}} & \multicolumn{1}{c}{\textbf{Acc}} & \multicolumn{1}{c}{\textbf{F1}} & \multicolumn{1}{c}{\textbf{Acc}} & \multicolumn{1}{c}{\textbf{F1}} & \multicolumn{1}{c}{\textbf{Acc}} & \multicolumn{1}{c}{\textbf{F1}}\\
\midrule
\multicolumn{18}{c}{\textbf{\textit{Macro Study: Transformation of the Feature Map Number}}} \\
\midrule  
\multirow{1}{*}{Leap Approach} 
  & \multicolumn{1}{c}{3996} 
  & \multicolumn{1}{c}{{.635}} 
  & \multicolumn{1}{c}{{.630}} 
  & \multicolumn{1}{c}{{\textbf{.064}}} 
  & \multicolumn{1}{c}{{.058}} 
  & \multicolumn{1}{c}{{\textbf{.216}}}
  & \multicolumn{1}{c}{{\textbf{.206}}}
  & \multicolumn{1}{c}{.857} 
  & \multicolumn{1}{c}{.856} 
  & \multicolumn{1}{c}{{.980}} 
  & \multicolumn{1}{c}{{.980}} 
  & \multicolumn{1}{c}{{.976}} 
  & \multicolumn{1}{c}{{.976}} 
  & \multicolumn{1}{c}{{.126}} 
  & \multicolumn{1}{c}{{.109}} 
  & \multicolumn{1}{c}{{\textbf{.115}}} 
  & \multicolumn{1}{c}{{.093}} \\ 
  
\multirow{1}{*}{\textbf{Step Approach}}
  & \multicolumn{1}{c}{\textbf{1287}}
  & \multicolumn{1}{c}{{\textbf{.647}}} 
  & \multicolumn{1}{c}{{\textbf{.641}}} 
  & \multicolumn{1}{c}{{.063}} 
  & \multicolumn{1}{c}{{\textbf{.058}}} 
  & \multicolumn{1}{c}{{.184}} 
  & \multicolumn{1}{c}{{.177}} 
  & \multicolumn{1}{c}{{\textbf{.886}}} 
  & \multicolumn{1}{c}{{\textbf{.885}}} 
  & \multicolumn{1}{c}{{\textbf{.981}}} 
  & \multicolumn{1}{c}{{\textbf{.980}}} 
  & \multicolumn{1}{c}{{\textbf{.982}}} 
  & \multicolumn{1}{c}{{\textbf{.982}}} 
  & \multicolumn{1}{c}{{\textbf{.127}}} 
  & \multicolumn{1}{c}{{\textbf{.113}}} 
  & \multicolumn{1}{c}{{.114}} 
  & \multicolumn{1}{c}{{\textbf{.095}}} \\ 
  
\midrule
\multicolumn{18}{c}{\textbf{\textit{Macro Study: Transformation of the Feature Map Size}}} \\
\midrule
\multirow{1}{*}{Pyramid Approach} 
  & \multicolumn{1}{c}{1287} 
  & \multicolumn{1}{c}{{.647}} 
  & \multicolumn{1}{c}{{.641}} 
  & \multicolumn{1}{c}{{\textbf{.063}}}
  & \multicolumn{1}{c}{{\textbf{.058}}}
  & \multicolumn{1}{c}{{.184}} 
  & \multicolumn{1}{c}{{.177}} 
  & \multicolumn{1}{c}{{.886}} 
  & \multicolumn{1}{c}{{.885}} 
  & \multicolumn{1}{c}{{\textbf{.981}}}
  & \multicolumn{1}{c}{{\textbf{.980}}}
  & \multicolumn{1}{c}{{.982}} 
  & \multicolumn{1}{c}{{.982}} 
  & \multicolumn{1}{c}{{.127}} 
  & \multicolumn{1}{c}{{.113}} 
  & \multicolumn{1}{c}{{.114}} 
  & \multicolumn{1}{c}{{.095}} \\ 
  
\multirow{1}{*}{\textbf{Pagoda Approach}}
  & \multicolumn{1}{c}{\textbf{758}}
  & \multicolumn{1}{c}{{\textbf{.701}}} 
  & \multicolumn{1}{c}{{\textbf{.698}}} 
  & \multicolumn{1}{c}{.059}
  & \multicolumn{1}{c}{.053}
  & \multicolumn{1}{c}{{\textbf{.204}}} 
  & \multicolumn{1}{c}{{\textbf{.197}}}
  & \multicolumn{1}{c}{{\textbf{.911}}}
  & \multicolumn{1}{c}{{\textbf{.910}}}  
  & \multicolumn{1}{c}{.973}
  & \multicolumn{1}{c}{.973}
  & \multicolumn{1}{c}{{\textbf{.986}}} 
  & \multicolumn{1}{c}{{\textbf{.985}}} 
  & \multicolumn{1}{c}{{\textbf{.128}}} 
  & \multicolumn{1}{c}{{\textbf{.116}}}
  & \multicolumn{1}{c}{{\textbf{.116}}} 
  & \multicolumn{1}{c}{{\textbf{.096}}} \\ 
\midrule
\multicolumn{18}{c}{\textbf{\textit{Micro Study}}} \\
\midrule
\multirow{1}{*}{Vanilla Conv} 
  & \multicolumn{1}{c}{758} 
  & \multicolumn{1}{c}{.701}
  & \multicolumn{1}{c}{.698}
  & \multicolumn{1}{c}{.059}
  & \multicolumn{1}{c}{.053}
  & \multicolumn{1}{c}{.204}
  & \multicolumn{1}{c}{.197}
  & \multicolumn{1}{c}{.911}
  & \multicolumn{1}{c}{.910}
  & \multicolumn{1}{c}{.973}
  & \multicolumn{1}{c}{.973}
  & \multicolumn{1}{c}{.986}
  & \multicolumn{1}{c}{.985}
  & \multicolumn{1}{c}{.128}
  & \multicolumn{1}{c}{{\textbf{.116}}}
  & \multicolumn{1}{c}{.116}
  & \multicolumn{1}{c}{.096} \\ 
\multirow{1}{*}{Separable Conv} 
  & \multicolumn{1}{c}{492} 
  & \multicolumn{1}{c}{.704}
  & \multicolumn{1}{c}{.699}
  & \multicolumn{1}{c}{.061}
  & \multicolumn{1}{c}{.051}
  & \multicolumn{1}{c}{.181}
  & \multicolumn{1}{c}{.176}
  & \multicolumn{1}{c}{.915}
  & \multicolumn{1}{c}{.915}
  & \multicolumn{1}{c}{.979}
  & \multicolumn{1}{c}{.979}
  & \multicolumn{1}{c}{.985}
  & \multicolumn{1}{c}{.985}
  & \multicolumn{1}{c}{.118}
  & \multicolumn{1}{c}{.107}
  & \multicolumn{1}{c}{.111}
  & \multicolumn{1}{c}{.098} \\ 
\multirow{1}{*}{\textbf{Group Conv}}
  & \multicolumn{1}{c}{\textbf{482}} 
  & \multicolumn{1}{c}{{\textbf{.707}}}
  & \multicolumn{1}{c}{{\textbf{.704}}}
  & \multicolumn{1}{c}{{\textbf{.069}}}
  & \multicolumn{1}{c}{{\textbf{.061}}}
  & \multicolumn{1}{c}{{\textbf{.207}}}
  & \multicolumn{1}{c}{{\textbf{.200}}}
  & \multicolumn{1}{c}{{\textbf{.920}}}
  & \multicolumn{1}{c}{{\textbf{.920}}}
  & \multicolumn{1}{c}{{\textbf{.983}}}
  & \multicolumn{1}{c}{{\textbf{.983}}}
  & \multicolumn{1}{c}{{\textbf{.990}}}
  & \multicolumn{1}{c}{{\textbf{.990}}}
  & \multicolumn{1}{c}{{\textbf{.129}}}
  & \multicolumn{1}{c}{.112}
  & \multicolumn{1}{c}{{\textbf{.120}}}
  & \multicolumn{1}{c}{{\textbf{.103}}} \\ 
\bottomrule
\end{tabular}}
\end{table}
\endgroup

%% file: text/03-experiments.tex
\section{Experiments}\label{Experiments}

\input{./tables/Experiment_table}

To assess the practical value of the recipe, we evaluate NeuroSketch-Base and NeuroSketch-Large through more than 3,000 experiments. Section~\ref{Experiment Setup Sec3} describes the evaluation setup, and Section~\ref{Results Sec 3} reports results for speech, visual, and auditory decoding. Section~\ref{Neurophysiological Interpretability} presents a case study of the spatiotemporal patterns captured by NeuroSketch-Large.

\subsection{Experimental Setup}\label{Experiment Setup Sec3}
The evaluation covers the eight tasks introduced in Section~\ref{roadmap} and includes NeuroSketch-Base (1.4M parameters) and NeuroSketch-Large (4.2M parameters). Appendix~\ref{NeuroSketch Implementation} provides their architectures and implementation details. The comparison includes ten baselines spanning time-series modeling, computer vision, and neural decoding. ModernTCN~\citep{Luo_2024_Moderntcn} and MedFormer~\citep{Wang_2024_Medformer} represent time-series models, while ConvFormer and CAFormer~\citep{Yu_2023_Metaformer} serve as vision backbones. Neural decoding baselines comprise DeepConvNet~\citep{schirrmeister2017deep}, EEGNet~\citep{Lawhern_2018_EEGNet}, EEG-Conformer~\citep{Song_2022_EEG}, and SPaRCNet~\citep{Jing_2023_Development}. We also evaluate the iEEG foundation model seegnificant~\citep{mentzelopoulos2024neural} and the EEG foundation model CBraMod~\citep{Wang_2025_CBraMod}. Both encoders are initialized with official pretrained weights and jointly fine-tuned with newly initialized classification heads. Further details on the baselines and evaluation protocol are provided in Appendices~\ref{Baseline Models} and~\ref{Experimental Settings}, respectively.

\subsection{Results}\label{Results Sec 3}

\vpara{Main results.}As shown in Table~\ref{tab:Table.4}, NeuroSketch outperforms all selected baselines, underscoring its strong capacity to model brain signals across diverse decoding scenarios. In addition, we note that NeuroSketch-Base performs similarly to NeuroSketch-Large on all datasets except Du-IN, where NeuroSketch-Large outperforms NeuroSketch-Base by a substantial margin. We further examine this scaling phenomenon in Appendix~\ref{Scaling Behavior}. To investigate the empirical performance gap between CNN-based and Transformer-based architectures, we analyze their representations in Appendix~\ref{Representation Analysis of CNN-2D and Transformer‑Based Architectures}. Complementary analyses examine kernel size and convolution group count (Appendix~\ref{Hyperparameter Analysis}), as well as pooling choices and residual connections (Appendix~\ref{Component-wise Ablation Study}). Further evaluations assess generalization to unseen subjects (Appendix~\ref{Cross-Subject Evaluation}), robustness to reduced training data (Appendix~\ref{Robustness to Reduced Training Data}), and training and inference efficiency (Appendix~\ref{Computational Efficiency}).

\vpara{Speech decoding.}On Du-IN, NeuroSketch-Large achieves 65.2\% accuracy, compared with 39.4\% for the strongest baseline, ConvFormer. By comparison, the iEEG foundation model seegnificant reaches 5.3\%, which may reflect the limited capacity of its single-layer Transformer architecture on this task. For semantic classification during sentence reading and imagination, our model instantiations also lead on both Chisco-R and Chisco-I despite the low overall accuracies. On Chisco-R, the Large variant attains 11.6\% accuracy versus 9.5\% for EEG-Conformer, a relative improvement of 22.1\%. On Chisco-I, the Base variant reaches 10.3\%, exceeding EEG-Conformer's 8.7\%.

% \vpara{Visual decoding.} As shown in Table{~\ref{tab:Table.4}}, for static image decoding on the FacesHouses dataset, NeuroSketch outperforms the second-best performing model, ModernTCN, with a 4.0\% improvement in accuracy. Furthermore, on the considerably more challenging ThingsEEG dataset, NeuroSketch achieves competitive performance with the state-of-the-art brain foundation model, CBraMod, while boasting a 1.5× higher inference throughput. ConvFormer and CAFormer also demonstrate strong visual decoding performance on the FacesHouses and ThingsEEG datasets, further suggesting the effectiveness of the CNN-2D architecture for neural decoding. For the highly challenging video decoding task, where most baseline models perform near the chance level (2.5\% accuracy), NeuroSketch demonstrates a clear advantage with an accuracy of 4.9\%. This result highlights its capacity to capture complex neural representations associated with dynamic visual stimuli.

% \vpara{Auditory decoding.} The results in Table{~\ref{tab:Table.4}} show that NeuroSketch attains state-of-the-art performance on both auditory decoding tasks, surpassing the ModernTCN and SPaRCNet models by 3.1\% and 3.9\% in accuracy, respectively. At the same time, several baseline models exhibit poor performance in this setting, with accuracy nearing chance level. These results highlight the generality and robustness of our proposed framework for auditory decoding applications.

\vpara{Visual decoding.}For static image decoding on FacesHouses, NeuroSketch-Large achieves 87.9\% accuracy, compared with 83.5\% for the strongest baseline, ModernTCN. This advantage also holds on the more challenging ThingsEEG dataset, where NeuroSketch-Base reaches 19.6\%, exceeding SPaRCNet's 18.9\%. For video decoding on SEED-DV, NeuroSketch-Large achieves 4.7\% accuracy, while most baselines perform near chance level (2.5\%).

\vpara{Auditory decoding.}NeuroSketch achieves the best or tied-best accuracy among the evaluated models on both auditory perception and imagery tasks. For perception (OpenMIIR-P), NeuroSketch-Large reaches 98.2\%, exceeding the strongest brain-domain baseline, CBraMod (97.5\%). For imagery (OpenMIIR-I), its accuracy of 97.7\% matches CAFormer and slightly exceeds ConvFormer (97.6\%). This comparable imagery performance is achieved with approximately one-fifth as many parameters as the two vision baselines.

\subsection{Case Study}\label{Neurophysiological Interpretability}

% \begin{figure}[t]
%    \centering
%    \includegraphics[width=0.90\linewidth]{./figures/Fig5.pdf}
%    \caption{\textbf{Score‑CAM saliency maps for Subjects 02 and 11 in the Du‑IN dataset.} Warmer colors represent higher feature importance to model predictions. The visualization reveals two key localization patterns: (1) \textbf{spatial locality}—activations concentrate on the ventral sensorimotor cortex (vSMC) and bilateral superior temporal gyrus (STG), aligning with established speech motor regions; (2) \textbf{temporal locality}—activations peak during the mid‑trial period (1200--1800\,ms for Subject 02; 800--1600\,ms for Subject 11), with minimal contributions from early and late segments.}
%    \label{fig:Fig5}
% \end{figure}

\begin{figure}[t]
   \centering
   \includegraphics[width=1.0\linewidth]{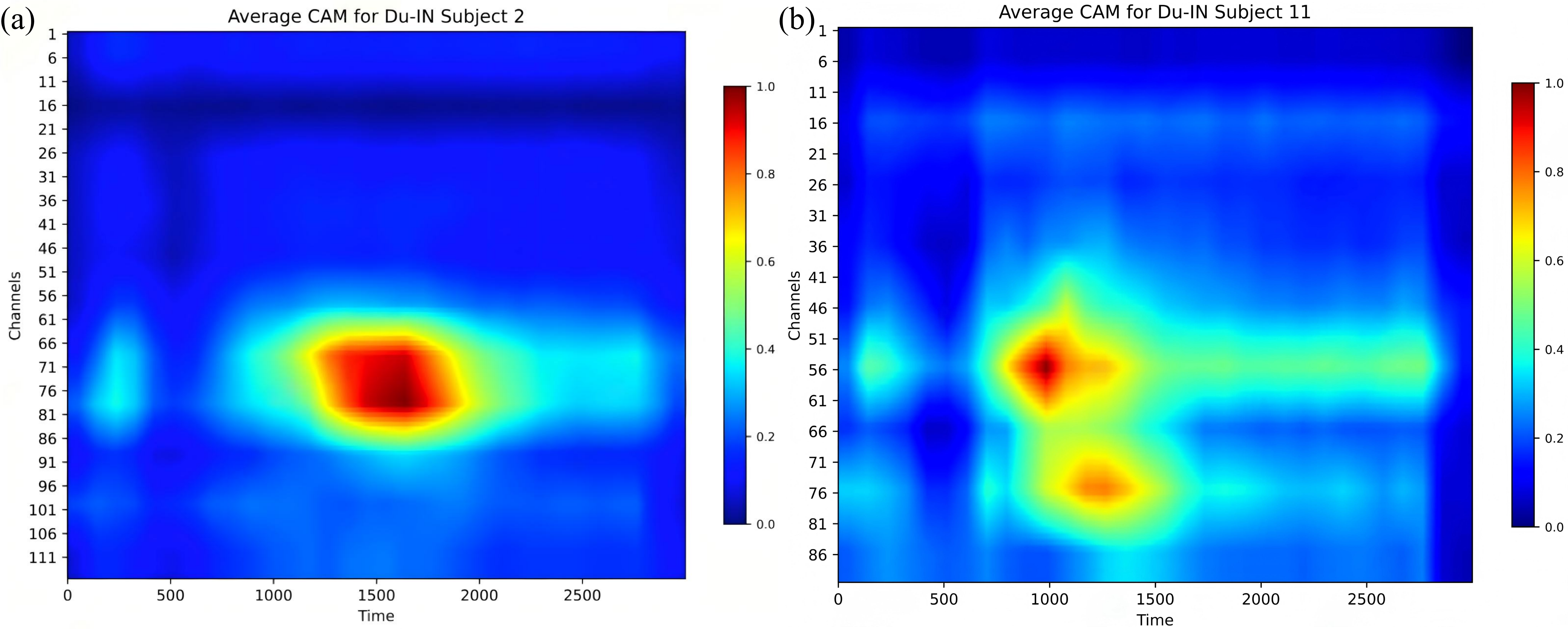}
  \caption{\textbf{Score-CAM saliency maps on Du-IN.}
  \textbf{(a)} Subject 02. \textbf{(b)} Subject 11.
  Maps are averaged across samples for each subject. Warmer colors indicate greater importance to model predictions. Saliency is concentrated in specific cortical regions and mid-trial time windows, highlighting spatial and temporal locality.}
  
   \label{fig:Fig5}
\end{figure}

To examine the spatiotemporal features captured by NeuroSketch, we conducted a case study using Subjects 02 and 11 from Du-IN. For this analysis, NeuroSketch-Large was trained using all 115 channels from Subject 02 and all 90 channels from Subject 11. Score-CAM~\citep{wang2020score} was then applied to the trained model to visualize the spatial and temporal regions associated with its predictions.

Score-CAM estimates feature-map importance without gradients and projects the resulting activations onto the input space. This procedure yields a saliency map for each sample, highlighting the spatiotemporal regions associated with the predicted class. Averaging these maps across samples from each subject produces the subject-level maps shown in Figure~\ref{fig:Fig5}. These aggregated maps provide the basis for identifying spatial and temporal regions of interest (ROIs).

To characterize the spatial patterns in these maps, we averaged each subject-level map over time to obtain a saliency score for each channel. The channels with the highest scores cluster around the ventral sensorimotor cortex (vSMC) and bilateral superior temporal gyrus (STG). These regions have established roles in speech processing, with vSMC coordinating articulatory movements and STG processing auditory feedback~\citep{bouchard2013functional,hickok2007cortical,chartier2018encoding}. This correspondence supports the neurophysiological interpretability of the spatial patterns captured by NeuroSketch.
Along the temporal dimension, the saliency maps show peak activations during the middle of each trial, with weaker contributions from earlier and later segments. These peaks occur at 800--1600\,ms for Subject 11 and 1200--1800\,ms for Subject 02. Such temporal concentration is consistent with the transient nature of task-relevant neural activity.

Together, these observations are consistent with the temporal and spatial motivations underlying the practical design recipe.

%% file: tables/Experiment_table.tex
\begingroup
\setlength{\tabcolsep}{1.0pt}   % Set the uniform column spacing
\renewcommand{\arraystretch}{1.0} % Reduce row height
\newcommand{\var}[1]{{\fontsize{6}{7}\selectfont #1}}
\setlength{\heavyrulewidth}{1.0pt}   % Top and bottom line thickness
\setlength{\lightrulewidth}{0.5pt}    % Middle line thickness
\setlength{\arrayrulewidth}{0.5pt}    % Uniform thickness of horizontal lines

\begin{table}[t]
\centering % 让表格居中

\caption{\textbf{Average classification accuracy} (mean $\pm$ std across three folds) for models on multiple neural decoding datasets. The best results are in \textbf{bold}. We use \textemdash\ to denote entries that are not applicable or not evaluated. Due to substantial differences between iEEG and EEG, we evaluate each pretrained model only on the modality it was pretrained on, the corresponding cross-modality cells are therefore marked with \textemdash. In addition, CBraMod uses a patch size of 200, whereas \emph{ThingsEEG} has an input length of 100, so CBraMod cannot be run on \emph{ThingsEEG} and is marked \textemdash.}

\label{tab:Table.4}
\NeuroSketchFitTable{ % 调整表格宽度为文本宽度，高度自适应
\fontsize{8}{9.5}\selectfont
\begin{tabular}{@{} p{1.2cm} p{0.55cm} p{0.55cm} p{0.55cm} p{0.55cm} p{0.55cm} p{0.55cm} p{0.55cm} p{0.55cm} p{0.55cm} p{0.55cm} p{0.55cm} p{0.55cm} p{0.55cm} p{0.55cm} p{0.55cm} p{0.55cm} p{0.55cm} @{}}
\toprule
% \multicolumn{2}{c}{\multirow{3}{*}{\textbf{Model}}} &
\multicolumn{2}{c}{\multirow{3}{*}{\textbf{\diagbox{Models}{Datasets}}}} & 
\multicolumn{6}{c}{\textbf{Speech Decoding}} & 
\multicolumn{6}{c}{\textbf{Visual Decoding}} & 
\multicolumn{4}{c}{\textbf{Auditory Decoding}} \\
\cmidrule(lr){3-8} \cmidrule(lr){9-14} \cmidrule(lr){15-18} 
& \multicolumn{1}{c}{}
& \multicolumn{2}{c}{\multirow{2}{*}{\textbf{Du-IN}}}
& \multicolumn{2}{c}{\multirow{2}{*}{\textbf{\makecell{Chisco-R}}}} 
& \multicolumn{2}{c}{\multirow{2}{*}{\textbf{\makecell{Chisco-I}}}}
& \multicolumn{2}{c}{\multirow{2}{*}{\textbf{SEED-DV}}} 
& \multicolumn{2}{c}{\multirow{2}{*}{\textbf{\makecell{Things-\\EEG}}}} 
& \multicolumn{2}{c}{\multirow{2}{*}{\textbf{\makecell{Faces-\\Houses}}}} 
& \multicolumn{2}{c}{\multirow{2}{*}{\textbf{\makecell{Open-\\MIIR-P}}}} 
& \multicolumn{2}{c}{\multirow{2}{*}{\textbf{\makecell{Open-\\MIIR-I}}}} \\

% \multicolumn{2}{c}{\multirow{2}{*}{\textbf{Model}}} &
% % \multicolumn{1}{c}{\multirow{2}{*}{\textbf{\makecell{Params\\(M)}}}} & 
% \multicolumn{1}{c}{\multirow{2}{*}{\textbf{\makecell{Throughput\\(sample/s)}}}}
% % & \multicolumn{1}{c}{\multirow{2}{*}{\textbf{\makecell{FLOPs\\(G)}}}} 
% & \multicolumn{2}{c}{\multirow{2}{*}{\textbf{Du-IN}}}
% & \multicolumn{2}{c}{\multirow{2}{*}{\textbf{\makecell{Chisco-R}}}} 
% & \multicolumn{2}{c}{\multirow{2}{*}{\textbf{\makecell{Chisco-I}}}}
% & \multicolumn{2}{c}{\multirow{2}{*}{\textbf{SEED-DV}}} 
% & \multicolumn{2}{c}{\multirow{2}{*}{\textbf{\makecell{Things-\\EEG}}}} 
% & \multicolumn{2}{c}{\multirow{2}{*}{\textbf{\makecell{Faces-\\Houses}}}} 
% & \multicolumn{2}{c}{\multirow{2}{*}{\textbf{\makecell{Open-\\MIIR-P}}}} 
% & \multicolumn{2}{c}{\multirow{2}{*}{\textbf{\makecell{Open-\\MIIR-I}}}} \\

\\ 
\midrule
\multicolumn{2}{l}{\multirow{1}{*}{ConvFormer}} 
  % & \multicolumn{1}{c}{24} 
  % & \multicolumn{1}{c}{357}
  & \multicolumn{2}{c}{.394\var{$\pm$.005}} 
  & \multicolumn{2}{c}{.086\var{$\pm$.001}} 
  & \multicolumn{2}{c}{.080\var{$\pm$.004}} 
  & \multicolumn{2}{c}{.024\var{$\pm$.000}} 
  & \multicolumn{2}{c}{.151\var{$\pm$.003}} 
  & \multicolumn{2}{c}{.820\var{$\pm$.013}} 
  & \multicolumn{2}{c}{.975\var{$\pm$.005}} 
  & \multicolumn{2}{c}{.976\var{$\pm$.002}} 
\\ 
\multicolumn{2}{l}{\multirow{1}{*}{CAFormer}} 
  % & \multicolumn{1}{c}{26} 
  % & \multicolumn{1}{c}{367}
  & \multicolumn{2}{c}{.221\var{$\pm$.003}} 
  & \multicolumn{2}{c}{.081\var{$\pm$.003}} 
  & \multicolumn{2}{c}{.076\var{$\pm$.002}} 
  & \multicolumn{2}{c}{.025\var{$\pm$.001}} 
  & \multicolumn{2}{c}{.160\var{$\pm$.009}} 
  & \multicolumn{2}{c}{.800\var{$\pm$.012}} 
  & \multicolumn{2}{c}{.976\var{$\pm$.006}} 
  & \multicolumn{2}{c}{.977\var{$\pm$.002}} 
\\

\midrule
\multicolumn{2}{l}{\multirow{1}{*}{ModernTCN}} 
  % & \multicolumn{1}{c}{34} 
  % & \multicolumn{1}{c}{361}
  & \multicolumn{2}{c}{.339\var{$\pm$.005}} 
  & \multicolumn{2}{c}{.079\var{$\pm$.003}}
  & \multicolumn{2}{c}{.074\var{$\pm$.002}} 
  & \multicolumn{2}{c}{.040\var{$\pm$.001}} 
  & \multicolumn{2}{c}{.123\var{$\pm$.003}} 
  & \multicolumn{2}{c}{.835\var{$\pm$.006}}
  & \multicolumn{2}{c}{.964\var{$\pm$.010}}
  & \multicolumn{2}{c}{.935\var{$\pm$.002}} 
\\ 

\multicolumn{2}{l}{\multirow{1}{*}{MedFormer}} 
  % & \multicolumn{1}{c}{36} 
  % & \multicolumn{1}{c}{272}
  & \multicolumn{2}{c}{.212\var{$\pm$.006}} 
  & \multicolumn{2}{c}{.089\var{$\pm$.003}} 
  & \multicolumn{2}{c}{.085\var{$\pm$.001}} 
  & \multicolumn{2}{c}{.025\var{$\pm$.000}} 
  & \multicolumn{2}{c}{.045\var{$\pm$.002}} 
  & \multicolumn{2}{c}{.756\var{$\pm$.011}} 
  & \multicolumn{2}{c}{.948\var{$\pm$.001}} 
  & \multicolumn{2}{c}{.941\var{$\pm$.001}} 
  
\\ 

\midrule
\multicolumn{2}{l}{\multirow{1}{*}{DeepConvNet}} 
  % & \multicolumn{1}{c}{34} 
  % & \multicolumn{1}{c}{361}
  & \multicolumn{2}{c}{.053\var{$\pm$.002}} 
  & \multicolumn{2}{c}{.053\var{$\pm$.004}}
  & \multicolumn{2}{c}{.052\var{$\pm$.001}} 
  & \multicolumn{2}{c}{.029\var{$\pm$.001}} 
  & \multicolumn{2}{c}{.091\var{$\pm$.001}} 
  & \multicolumn{2}{c}{.822\var{$\pm$.005}}
  & \multicolumn{2}{c}{.896\var{$\pm$.009}}
  & \multicolumn{2}{c}{.789\var{$\pm$.005}} 

\\ 

\multicolumn{2}{l}{\multirow{1}{*}{EEGNet}} 
  % & \multicolumn{1}{c}{36} 
  % & \multicolumn{1}{c}{272}
  & \multicolumn{2}{c}{.060\var{$\pm$.003}} 
  & \multicolumn{2}{c}{.071\var{$\pm$.002}} 
  & \multicolumn{2}{c}{.082\var{$\pm$.002}} 
  & \multicolumn{2}{c}{.046\var{$\pm$.002}} 
  & \multicolumn{2}{c}{.070\var{$\pm$.000}} 
  & \multicolumn{2}{c}{.830\var{$\pm$.006}} 
  & \multicolumn{2}{c}{.915\var{$\pm$.010}} 
  & \multicolumn{2}{c}{.852\var{$\pm$.006}} 
\\ 
\multicolumn{2}{l}{\multirow{1}{*}{EEG-Conformer}} 
  % & \multicolumn{1}{c}{38} 
  % & \multicolumn{1}{c}{432}
  & \multicolumn{2}{c}{.205\var{$\pm$.003}} 
  & \multicolumn{2}{c}{.095\var{$\pm$.002}} 
  & \multicolumn{2}{c}{.087\var{$\pm$.002}}
  & \multicolumn{2}{c}{.033\var{$\pm$.002}} 
  & \multicolumn{2}{c}{.077\var{$\pm$.002}} 
  & \multicolumn{2}{c}{.762\var{$\pm$.001}} 
  & \multicolumn{2}{c}{.962\var{$\pm$.013}}
  & \multicolumn{2}{c}{.930\var{$\pm$.002}} 

\\ 
\multicolumn{2}{l}{\multirow{1}{*}{SPaRCNet}} 
  % & \multicolumn{1}{c}{24} 
  % & \multicolumn{1}{c}{152}
  & \multicolumn{2}{c}{{.026\var{$\pm$.001}}} 
  & \multicolumn{2}{c}{.005\var{$\pm$.000}} 
  & \multicolumn{2}{c}{.005\var{$\pm$.000}} 
  & \multicolumn{2}{c}{.026\var{$\pm$.001}}
  & \multicolumn{2}{c}{.189\var{$\pm$.001}} 
  & \multicolumn{2}{c}{.805\var{$\pm$.007}} 
  & \multicolumn{2}{c}{.969\var{$\pm$.007}} 
  & \multicolumn{2}{c}{{{.948\var{$\pm$.005}}}} 

\\ 
\midrule
\multicolumn{2}{l}{\multirow{1}{*}{seegnificant}} 
  % & \multicolumn{1}{c}{38} 
  % & \multicolumn{1}{c}{432}
  & \multicolumn{2}{c}{.053\var{$\pm$.002}} 
  & \multicolumn{2}{c}{\textemdash} 
  & \multicolumn{2}{c}{\textemdash} 
  & \multicolumn{2}{c}{\textemdash} 
  & \multicolumn{2}{c}{\textemdash} 
  & \multicolumn{2}{c}{.831\var{$\pm$.006}} 
  & \multicolumn{2}{c}{\textemdash}
  & \multicolumn{2}{c}{\textemdash} 
% \\ 
% \multicolumn{2}{c}{\multirow{1}{*}{EEGPT}} 
%   % & \multicolumn{1}{c}{34} 
%   & \multicolumn{1}{c}{437}
%   & \multicolumn{2}{c}{-} 
%   & \multicolumn{2}{c}{.005\var{$\pm$.000}} 
%   & \multicolumn{2}{c}{.005\var{$\pm$.000}} 
%   & \multicolumn{2}{c}{.025\var{$\pm$.000}} 
%   & \multicolumn{2}{c}{.083\var{$\pm$.001}} 
%   & \multicolumn{2}{c}{-} 
%   & \multicolumn{2}{c}{.682\var{$\pm$.006}} 
%   & \multicolumn{2}{c}{.352\var{$\pm$.009}} 
\\ 
\multicolumn{2}{l}{\multirow{1}{*}{CBraMod}} 
  % & \multicolumn{1}{c}{4} 
  % & \multicolumn{1}{c}{347}
  & \multicolumn{2}{c}{\textemdash} 
  & \multicolumn{2}{c}{.084\var{$\pm$.002}} 
  & \multicolumn{2}{c}{.077\var{$\pm$.001}} 
  & \multicolumn{2}{c}{.026\var{$\pm$.002}} 
  & \multicolumn{2}{c}{\textemdash} 
  & \multicolumn{2}{c}{\textemdash} 
  & \multicolumn{2}{c}{.975\var{$\pm$.011}} 
  & \multicolumn{2}{c}{.933\var{$\pm$.008}} 
\\ 
\midrule
\multicolumn{2}{l}{\multirow{1}{*}{NeuroSketch-Base}}
  % & \multicolumn{1}{c}{9} 
  % & \multicolumn{1}{c}{{{532}}}
  & \multicolumn{2}{c}{{{.472\var{$\pm$.006}}}} 
  & \multicolumn{2}{c}{{{.111\var{$\pm$.001}}}} 
  & \multicolumn{2}{c}{{\textbf{.103\var{$\pm$.002}}}} 
  & \multicolumn{2}{c}{{{.045\var{$\pm$.002}}}} 
  & \multicolumn{2}{c}{{\textbf{.196\var{$\pm$.002}}}} 
  & \multicolumn{2}{c}{{{.879\var{$\pm$.008}}}} 
  & \multicolumn{2}{c}{{{.981\var{$\pm$.010}}}} 
  & \multicolumn{2}{c}{{{.970\var{$\pm$.002}}}} 
\\ 
\multicolumn{2}{l}{\multirow{1}{*}{NeuroSketch-Large}} 
  % & \multicolumn{1}{c}{4} 
  % & \multicolumn{1}{c}{347}
  & \multicolumn{2}{c}{\textbf{.652\var{$\pm$.002}}} 
  & \multicolumn{2}{c}{\textbf{.116\var{$\pm$.001}}}
  & \multicolumn{2}{c}{.095\var{$\pm$.005}} 
  & \multicolumn{2}{c}
  {{\textbf{.047\var{$\pm$.003}}}}
  & \multicolumn{2}{c}{{{.177\var{$\pm$.001}}}} 
  & \multicolumn{2}{c}
  {{\textbf{.879\var{$\pm$.007}}}}
  & \multicolumn{2}{c}
  {{\textbf{.982\var{$\pm$.009}}}}
  & \multicolumn{2}{c}
  {{\textbf{.977\var{$\pm$.002}}}}
\\ 
\bottomrule
\end{tabular}}
\end{table}
\endgroup

%% file: text/04-conclusion_and_future_work.tex
\section{Conclusion and Future Work}\label{Conclusion and Future Work}

\vpara{Conclusion.}In this paper, we develop NeuroSketch, a practical design recipe for neural decoding, through a basic architecture study followed by macro- and micro-level optimization. The basic architecture study favors CNN-2D for modeling short-range temporal dynamics and local cross-channel dependencies. Building on this backbone, macro-level studies of latent space transformations support gradual feature-map expansion and early downsampling, while micro-level studies favor grouped convolutions. To evaluate the resulting recipe, we assess NeuroSketch-Base and NeuroSketch-Large across eight tasks spanning visual, auditory, and speech modalities and EEG, SEEG, and ECoG signals. This evaluation supports the practical value of the recipe as a guide for neural decoding model design.

\vpara{Limitations and future work.}The current practical design recipe is developed and evaluated in neural decoding settings. Given the favorable scaling behavior observed in Appendix~\ref{Scaling Behavior}, we hypothesize that scaling both data and model capacity will yield further gains. Accordingly, we will explore large-scale pretraining strategies guided by the NeuroSketch recipe in the future. We also plan to extend the evaluation to a broader range of neural decoding tasks to assess the recipe’s generalizability.

%% file: text/00-statements.tex
\subsection*{AI use statement}
In preparing this manuscript, we made limited use of a large language model (LLM) to assist with polishing the writing for grammar and clarity. In addition, the LLM was occasionally employed for small-scale code completion, such as generating boilerplate functions or suggesting minor syntax corrections. The authors developed the core ideas and experimental designs, conducted the analyses, and verified the conclusions.

\subsection*{Ethics statement}\label{Ethics statement}  
All datasets used in this study are publicly available. The corresponding dataset links are provided in Appendix~\ref{Dataset Description}. Consequently, our work does not involve personally sensitive information and does not pose apparent ethical concerns.

\subsection*{Reproducibility statement}\label{Reproducibility statement}
We place great importance on ensuring the reproducibility of our work. To this end, details of the preprocessing of each dataset are provided in Appendix~\ref{Data Preprocessing}, detailed model implementation and configuration are described in Appendix~\ref{Details of Models}. Appendix~\ref{Experimental Settings} documents data splitting, augmentation, model and training hyperparameters, foundation-model fine-tuning, and computational resources. The complete experimental results are available at \url{https://github.com/Galaxy-Dawn/NeuroSketch_Results}. Our code and scripts are available at \url{https://github.com/Galaxy-Dawn/NeuroSketch}.

%% file: text/05-related_work.tex
\section{Related Work}\label{Related Work}
\vpara{Neural decoding.}Neural decoding infers sensory experiences and cognitive states from recorded brain activity. Auditory decoding includes perception, which concerns responses to external sounds~\citep{Stober_2015_music, Sonawane_2021_GuessTheMusic}, and imagery, which concerns internally generated sounds~\citep{Stober_2015_music, DeBorst_2016_Brain-based}. Speech decoding targets linguistic information at different levels, including syllables~\citep{Feng_2023_high-performance}, words~\citep{Zheng_2024_Du-IN}, and sentences~\citep{Zhang_2024_Chisco}. Visual decoding aims to infer visual stimuli from neural activity. Reconstruction methods recover static images~\citep{Gaziv_2022_Self-supervised, Ahmadieh_2024_Visual, Singh_2023_EEG2IMAGE} or dynamic videos~\citep{Wen_2018_Neural, Liu_2024_EEG2video}.

\vpara{Neural architecture design.}Recent advances in neural architecture design have diverged into two directions: automated neural architecture search (NAS) and manual design driven by experience.~\citet{Tan_2019_Efficientnet} utilize NAS to scale all dimensions of depth, width, and resolution through an effective compound coefficient, leading to a family of models known as EfficientNets.~\citet{Radosavovic_2020_Designing} further integrate NAS with manual design principles to explore the design space, resulting in the development of RegNet. Meanwhile, manual architectural innovations have reinvigorated traditional components.~\citet{Yu_2022_Metaformer} reveal that the general Transformer architecture, termed MetaFormer rather than any specific token mixer (e.g., attention), is essential for achieving high performance.~\citet{Liu_2022_convnet} modernize a standard CNN towards the design of Swin Transformer~\citep{Liu_2021_Swin} through macro- and micro-level optimizations, proposing a family of pure CNNs called ConvNeXt.

%% file: text/06-appendix.tex
\section{Details of Datasets}\label{Details of Datasets}

\subsection{Dataset Description}\label{Dataset Description}
The evaluation uses six datasets covering eight neural decoding tasks. The dataset descriptions below summarize their recording settings and experimental paradigms.

\textbf{Du-IN}~\citep{Zheng_2024_Du-IN} records SEEG signals during spoken Mandarin word production. Each trial displays a word in white for 0.5 seconds, followed by a 2-second green cue prompting articulation. The word then disappears and a fixation cross appears. The dataset includes 12 participants, with 7--13 SEEG electrodes providing 72--158 recording channels per participant. Each participant reads 61 Chinese words, repeated 50 times each, while SEEG and audio are recorded simultaneously. Each participant contributes approximately 15 hours of data sampled at 2000 Hz, including around 3 hours of task-related activity. These segments yield approximately 3000 three-second trials and are subsequently downsampled to 1000 Hz. The dataset is available under CC BY 4.0 at \url{https://huggingface.co/datasets/liulab-repository/Du-IN}.

\textbf{Chisco}~\citep{Zhang_2024_Chisco} records EEG during silent reading and imagined speech. Three healthy participants were recorded with a 125-channel system at 1000 Hz. The experiment comprises 45 blocks of 150 trials. Each trial includes 5 seconds of silent reading and 3.3 seconds of imagined speech, with continuous EEG recording. The stimuli comprise 6,681 Chinese sentences of 6--15 characters from 39 semantic categories. Ten trials per block are randomly selected for verbal recall. A recall differing from the original sentence by more than four characters is marked incorrect. Two or more errors trigger a rest period followed by repetition of the block. The dataset is available under CC0 at \url{https://openneuro.org/datasets/ds005170/versions/1.1.2}.

\textbf{FacesHouses}~\citep{Miller_2019_library} records ECoG during the viewing of static images. The dataset includes 14 epilepsy patients with subdural electrode strips over the inferior temporal cortex. Participants view grayscale faces and houses in randomized order while ECoG is recorded at 1000 Hz. Each experiment comprises three sessions, each containing 50 face images and 50 house images. Images appear for 400 ms, separated by 400-ms blank intervals. The dataset is available under CC BY-SA at \url{https://purl.stanford.edu/zk881ps0522}.

\textbf{ThingsEEG}~\citep{gifford2022large} records EEG responses to images presented in a rapid serial visual presentation (RSVP) paradigm. Ten healthy participants were recorded with a 64-channel system at 1000 Hz. Each training image is presented four times, and each test image 80 times. Stimuli appear at a 200-ms stimulus onset asynchrony, with a target-detection task used to maintain engagement. Our decoding evaluation uses the 200-image test partition, as detailed in Appendix~\ref{Data Preprocessing}. The dataset is available under CC BY 4.0 at \url{https://osf.io/hd6zk/}.

\textbf{SEED-DV}~\citep{Liu_2024_EEG2video} records EEG at 1000 Hz from 20 participants viewing video clips. The stimuli comprise 1,400 two-second clips covering 40 semantic concepts, with 35 clips per concept. Each participant completes seven blocks, separated by rest periods. Each block includes all 40 concepts in randomized order, with five clips per concept. Participants receive the target concept before viewing its associated clips. Access requires an application at \url{https://bcmi.sjtu.edu.cn/ApplicationForm/apply_form/}.

\textbf{OpenMIIR}~\citep{Stober_2015_music} records EEG during music perception and imagery. The dataset includes 10 participants, eight with formal musical training, recorded with a 64-channel system at 512 Hz. The experiment comprises two sessions of five trials each. Twelve musical stimuli are presented in randomized order under four conditions: perception with auditory cues, imagery with cues, imagery without cues, and imagery without cues with feedback. The stimuli use 3/4 or 4/4 time signatures and tempos of 104--212 beats per minute. They include songs with lyrics, corresponding instrumental versions, and instrumental pieces. The dataset is available under ODC-PDDL at \url{https://hyper.ai/en/datasets/5591}.

\subsection{Data Preprocessing}\label{Data Preprocessing}

The following procedures define the inputs and classification targets used in our evaluation, following the preprocessing described in the source studies.

For \textbf{Du-IN}, we select 10 SEEG channels per subject using the configuration from the original study. Signals are downsampled to 1000 Hz and segmented into 2.5-second inputs. Each input is assigned one of the 61 spoken-word labels.

For \textbf{Chisco}, the tasks are silent reading (Chisco-R) and imagined speech (Chisco-I). Three noisy channels are removed, leaving 122 EEG channels, and signals are downsampled to 500 Hz. Inputs span 5 seconds for Chisco-R and 3.3 seconds for Chisco-I. Both tasks use the 39 semantic categories as classification targets.

For \textbf{FacesHouses}, continuous ECoG recordings are segmented using stimulus markers, retaining only face and house trials. Channel-wise z-score normalization reduces differences in signal scale. Inputs contain 31--102 channels per subject, sampled at 1000 Hz over 400 ms. The classification target distinguishes faces from houses.

For \textbf{ThingsEEG}, we use the preprocessed test partition released by the dataset authors to construct a 200-class supervised decoding task. Each image has 80 repetitions, yielding a tensor of shape $200\times80\times17\times100$ for each subject. The 17 channels are O1, Oz, O2, PO7, PO3, POz, PO4, PO8, P7, P5, P3, P1, Pz, P2, P4, P6, and P8. These channels cover occipital and parietal recording locations. Reshaping the first two axes yields 16,000 labeled trials, each containing 17 channels and 100 time steps. For the main evaluation, 20\% of trials form a fixed test set; the remaining 80\% are used for three-fold training and validation. Models are evaluated separately for each subject. This setting measures classification among the 200 selected images rather than generalization to unseen image classes.

For the \textbf{SEED-DV} dataset, we adopted the first benchmark task from the original SEED-DV study: 40-class classification of fine-grained video concepts, to evaluate dynamic visual stimulus decoding. Following the preprocessing protocol described in the original work, we downsampled the EEG signals to 200 Hz, with each trial spanning 2 seconds and containing 62 channels. The classification labels correspond to the 40 predefined video concepts. To standardize our evaluation pipeline, we merged data across all blocks for each subject, randomly shuffled the trials, and partitioned them into training, validation, and test sets.

For \textbf{OpenMIIR}, we evaluate perception (OpenMIIR-P) and imagery (OpenMIIR-I) using nine participants: P01, P04, P06, P07, P09, P11, P12, P13, and P14. EEG remains at its original sampling rate of 512 Hz and is band-pass filtered from 0.5 to 30 Hz. Epochs are extracted relative to audio-onset events. Perception uses condition 1, whereas imagery pools conditions 2--4. Each trial is divided into non-overlapping 600-sample windows across 64 channels, and each window is assigned one of the 12 musical-stimulus labels. Channel-wise z-score normalization uses the mean and standard deviation calculated across all extracted windows and time samples for each participant and task.

\section{Details of Models}\label{Details of Models}

\subsection{Models in the roadmap exploration} \label{Models in the roadmap exploration}
This section specifies the architectures used to develop the practical design recipe in Section~\ref{roadmap}.

\vpara{Basic architecture study.}The model depth and embedding dimension during the basic architecture study are shown in Table~\ref{tab:Table.5}. In both the CNN-1D and CNN-2D architectures, the convolution kernel size is set to 3. In CNN-1D, the number of channels increases from 1 to 64 in the stem layer, followed by 4 stages, each consisting of 5 layers. The number of channels in these stages increases progressively from 128 to 256, 512, and finally 1024. In CNN-2D, the channel configuration also starts from 1 to 64 in the stem layer. This is followed by 4 stages, each containing 4 layers, with the number of channels increasing from 96 to 192, 384, and 768, respectively. In the CNN-GRU architecture, the ratio of convolutional layers to GRU layers is set to 3:1. In the CNN-Transformer architecture, the initial ratio of convolutional layers to Transformer modules is set to 4:1. To fairly investigate how short- and long-term temporal information impacts model performance, we fix the total number of layers at 20 and systematically adjust the proportion of Transformer modules. To systematically evaluate the effectiveness of the proposed patching method, we vary only the data segmentation approach while keeping the embedding and subsequent self-attention-based feature extraction pipeline unchanged. All other hyperparameters are fixed to ensure that the observed performance differences can be attributed to the effectiveness of the patching method.

\input{./tables/Model_table}

\vpara{Macro optimization.}From a macro perspective, we optimize the transformation of the latent space during forward propagation. Specifically, we investigate two strategies for increasing the number of feature maps: the step approach and the leap approach. In the step approach, the number of feature maps is gradually increased: starting from 1 to 64 through the stem layers, then progressively to 96, 192, 384, and 768 across 4 stages, each containing 4 layers. In contrast, the leap approach increases the number of feature maps from 1 to 64 in the stem layer, and then directly increases it to 384 using an embedding layer. Subsequently, 16 layers are used to refine high-level features. We also investigate two strategies for decreasing the size of feature maps: the pyramid approach and the pagoda approach. In the pyramid approach, a downsampling layer is placed in each of the last three stages, with a downsampling ratio of 2 at each stage. In contrast, the pagoda approach places three identical downsampling layers consecutively within the second stage.

\vpara{Micro optimization.} From a micro perspective, we optimize the computation method. Specifically, we investigate two variants of vanilla convolution: the group convolution and the separable convolution. For the group convolution, the number of groups is set to 4. For the separable convolution, the number of groups is set to $C_{\mathrm{in}}$, followed by a pointwise convolution with a kernel size of 1 to aggregate channel-wise information.

\subsection{NeuroSketch Implementation} \label{NeuroSketch Implementation}
% detailed architecture and configuration of small and base nsk.
The NeuroSketch recipe is developed through staged architectural refinement and instantiated at two model scales. Starting from the approximately 35M-parameter CNN-2D backbone, macro-level optimization yields a 35.6M-parameter configuration. Grouped convolutions subsequently reduce the backbone parameter count to 14.4M. After establishing the recipe, we further reduce stage widths and depths to obtain its two instantiations, NeuroSketch-Base and NeuroSketch-Large, with approximately 1.4M and 4.2M parameters, respectively. The smaller models can still perform better because the gain mainly comes from a practical design recipe with a more suitable backbone family and improved macro and micro design choices, rather than from parameter count alone. Once the architecture is better aligned with the temporal-spatial structure of neural decoding, width and depth scaling can further improve parameter efficiency while preserving performance. Table~\ref{tab:Table.6} summarizes the configurations of both variants. Their shared architecture is described below.

Given an input $\mathbf{X}\in\mathbb{R}^{B\times C\times L}$, each model folds groups of three consecutive time samples into a channel-like dimension. This produces a two-dimensional representation $\mathbf{X}'\in\mathbb{R}^{B\times 1\times 3C\times \lceil L/3\rceil}$. The transformation is designed to expand the local temporal context available to the first convolutional layer while limiting early mixing across recording channels. Applied directly to the original $C\times L$ representation, a $3\times3$ kernel spans three adjacent recording channels but only three consecutive time samples. Folding the temporal axis allows the same kernel to cover a wider temporal range, supporting short-range temporal modeling before broader cross-channel integration in subsequent layers.

The reshaped two-dimensional representation is subsequently processed by a stem stage comprising four sequential stem blocks. Each block consists of a two-dimensional convolutional layer, followed by batch normalization and a ReLU activation function. The convolutional layers in all four blocks utilize a kernel size of 3, padding of 1, and strides of 2, 1, 1, and 2, respectively. The input and output channel dimensions for the four blocks are as follows: from 1 to 64, 64 to 32, 32 to 64, and 64 to 96, respectively.

Following the stem stage, there are four feature extraction stages. The input/output channel dimensions for these stages are [96 $\rightarrow$ $D_\text{stage 1}$], [$D_\text{stage 1}$ $\rightarrow$ $D_\text{stage 2}$], [$D_\text{stage 2}$ $\rightarrow$ $D_\text{stage 3}$], [$D_\text{stage 3}$ $\rightarrow$ $D_\text{stage 4}$]. Each stage contains $d$ blocks with two key components:

\begin{itemize}[leftmargin=*]
    \item [1.] \textbf{Patch Embedding:} In the second stage, every block uses a $3\times3$ convolution with stride 2 followed by batch normalization, resulting in two downsampling operations in NeuroSketch-Base and three in NeuroSketch-Large. In the other stages, a $3\times3$ convolution with stride 1 followed by batch normalization is used when the number of feature maps changes. Otherwise, an identity mapping is applied.
    \item [2.] \textbf{Convolution Module:} This component applies grouped $3\times 3$ convolutions, with the number of groups equal to $G$ to efficiently capture local channel-wise dependencies. It is followed by batch normalization, ReLU activation, and a $1\times 1$ convolution for feature fusion. The output is added back to the patch-embedded input via a residual connection.
\end{itemize}

After the four feature extraction stages, we obtain a tensor of shape $\mathbb{R}^{B\times D_\text{stage 4} \times C' \times T'}$, where $C'$ and $T'$ denote the downsampled channel and temporal dimensions. We then apply GeM pooling to aggregate it into a representation $\mathbb{R}^{B\times D_\text{stage 4}}$, which is finally passed through a linear layer to produce the class probabilities.

\input{./tables/nsk_details}

\subsection{Baseline Models} \label{Baseline Models}
Here, we introduce the details of the baselines for performance evaluation in Section~\ref{Experiments}.

\textbf{ConvFormer} and \textbf{CAFormer}~\citep{Yu_2023_Metaformer} are two variants of MetaFormer~\citep{Yu_2022_Metaformer}, employing different token mixers. ConvFormer uses depthwise separable convolution as its token mixer. With this design, ConvFormer can be regarded as a pure CNN model that does not rely on channel or spatial attention mechanisms. CAFormer uses depthwise separable convolution as the token mixer in the first two stages of the model and self-attention in the last two stages. This design enables CAFormer to capture local features while better obtaining long-range dependencies.

\textbf{ModernTCN}~\citep{Luo_2024_Moderntcn} is a modernized purely convolutional architecture. It enhances the traditional TCN by incorporating depthwise convolutions and convolutional feed-forward networks (ConvFFNs) into the 1D CNN design. Additionally, it introduces time series–specific adaptations, such as patchified variable-independent embeddings, to improve suitability for time series tasks.

\textbf{MedFormer}~\citep{Wang_2024_Medformer} is a multi-granularity patching Transformer specifically designed for medical time series classification. It effectively captures the features of medical time series through cross-channel patching, multi-granularity embedding, and a two-stage multi-granularity self-attention mechanism.

\textbf{DeepConvNet}~\citep{schirrmeister2017deep} is an EEG decoding CNN. Temporal convolution first extracts frequency-related features, followed by spatial convolution across recording channels. Subsequent convolution, batch normalization, ELU, and max-pooling layers learn hierarchical representations, with dropout used during training. A final dense layer performs classification.

\textbf{EEGNet}~\citep{Lawhern_2018_EEGNet} is a compact CNN tailored for EEG-based BCI. It decouples frequency and spatial filtering by using temporal convolutions as learnable band-pass filters followed by depthwise spatial convolutions across channels, and employs separable (pointwise) convolutions for efficient feature mixing. EEGNet attains strong performance across diverse BCI paradigms while using fewer than 0.5 million parameters, making it data-efficient and well suited to small EEG datasets.

\textbf{EEG-Conformer}~\citep{Song_2022_EEG} is a CNN-Transformer model for EEG signals. The convolution module learns low-level local features through one-dimensional temporal and spatial convolution layers, and the self-attention module processes the output of the convolution module to learn global temporal dependencies.

\textbf{SPaRCNet}~\citep{Jing_2023_Development} is a 1D CNN for seizure pattern recognition. Its structure features dense and transition blocks. Each dense block has four layers of two convolutional layers and two ELUs, and each transition block contains an ELU, a convolutional layer, and an average pooling layer.

\textbf{seegnificant}~\citep{mentzelopoulos2024neural} is a brain foundation model for cross-subject neural decoding from SEEG. It tokenizes electrode-wise signals using convolutions, models long-term temporal dependencies with self-attention, and integrates electrode 3D spatial locations through positional encoding followed by cross-electrode attention. A unified backbone extracts global neural representations, while subject-specific task heads enable individualized decoding. Trained on multi-session SEEG data from 21 participants, seegnificant demonstrates effective behavioral response time decoding and supports few-shot transfer to new subjects, offering a path toward robust multi-subject generalization in SEEG analysis.

\textbf{CBraMod}~\citep{Wang_2025_CBraMod} is a brain foundation model for EEG decoding. The model first segments EEG signals into patches and randomly masks them. After patch encoding and asymmetric conditional position encoding, it learns spatio-temporal dependencies through the parallel spatial and temporal attention mechanisms of the cross-Transformer. Finally, it uses the reconstruction head to reconstruct the masked EEG patches, thereby learning the general representation of EEG signals.

\section{Experimental Settings}\label{Experimental Settings}
This section describes the training and evaluation procedures used in Sections~\ref{roadmap} and~\ref{Experiments}.

\subsection{Data splitting}\label{Data splitting}
For Section~\ref{roadmap}, each subject's data is split into training, validation, and test sets with a 3:1:1 ratio. For Section~\ref{Experiments}, we allocate 20\% of the data for testing, and the remaining 80\% is used for a 3-fold cross-validation. Specifically, the remaining data are divided into three equal parts, with each iteration using one part as the validation set while the other two are combined for training, ensuring a comprehensive and reliable assessment of the model's effectiveness. 

\subsection{Data augmentation}\label{Data augmentation}
To further enhance data diversity, we employ a strong data augmentation strategy comprising the following techniques:

\vpara{Random Shift.}We define a maximum shift range proportional to the input sequence length. For each training instance, a shift step is randomly sampled from this range. A positive value indicates a forward shift of the sequence, whereas a negative value corresponds to a backward shift. This approach enhances the model's robustness to uncertainty in stimulus onset times. 

\vpara{Noise.}We generate noise from a standard normal distribution matching the shape of the original data. The noise is scaled by a predefined standard deviation and added to the input to produce noisy data. This method improves the model’s resilience to signal perturbations.

\vpara{Channel Masking.}During training, a mask is applied to each channel with the specified probability, zeroing out the corresponding channel values. This technique reduces over-reliance on specific channels and promotes better integration of multi-channel information.

\vpara{Time Masking.}Similar to channel masking, we apply masks along the temporal dimension. This encourages the model to extract features robustly across various time segments and enhances generalization.

\vpara{Mixup.}A mixing coefficient $\lambda$ is sampled from a Beta distribution parameterized by a hyperparameter $\alpha=0.4$. The original sample and a randomly chosen sample are linearly combined using $\lambda$, and their corresponding labels are mixed accordingly. This augmentation method exposes the model to a broader range of data combinations, thereby improving generalization.

\subsection{Model and Training Hyperparameters}\label{Hyperparameters}
The following descriptions summarize the configurations of the ten baselines. The shared training hyperparameters are provided in Table~\ref{tab:Table.7}. For reference, parameter counts are calculated using inputs with 64 recording channels and 2,000 time samples, with 12 output classes. Counts include all parameters registered in each model implementation, including classification heads. Under this setting, NeuroSketch-Base and NeuroSketch-Large contain approximately 1.4M and 4.2M parameters, respectively. Their stage widths and depths are specified in Table~\ref{tab:Table.6}.

ModernTCN uses three stages with 32, 64, and 128 feature channels, respectively, and one block per stage. The large temporal kernels have sizes 21, 19, and 13, while the small kernels have size 5 throughout. Each feed-forward module expands the feature width by a factor of four. The input patch length and stride are both 50 time samples. This configuration contains approximately 8.1M parameters.

MedFormer processes temporal patches of 5, 10, and 20 samples using 128-dimensional embeddings. Its encoder contains six layers, each with eight attention heads and a feed-forward hidden dimension of 256. This configuration contains approximately 2.3M parameters.

ConvFormer and CAFormer use their S18 backbones, with approximately 26.8M and 26.3M parameters, respectively. Both models use the same input preprocessing as NeuroSketch to construct a 2D representation of the neural recordings. GeM pooling converts the backbone outputs into 512-dimensional representations, which are passed to task-specific linear classifiers.

DeepConvNet uses four convolutional blocks with 25, 50, 100, and 200 feature channels. Temporal convolutions use kernels of length 5, and each block halves the temporal resolution through pooling. LeakyReLU activation and dropout with a rate of 0.5 are used throughout. This configuration contains approximately 0.5M parameters.

EEGNet starts with 16 temporal filters and learns two spatial filters for each temporal filter, producing 32 feature maps. Subsequent convolutional processing retains 32 feature maps, with dropout set to 0.25. The flattened features are passed to a linear classifier. This configuration contains approximately 1.5M parameters.

EEG-Conformer uses 40-dimensional embeddings and six Transformer encoder layers with five attention heads per layer. Its convolutional input module applies temporal pooling with a stride of 50 samples. The encoded features are averaged over time and passed through layer normalization and a linear classifier. This configuration contains approximately 0.2M parameters.

SPaRCNet starts with 64 feature channels and uses six dense blocks, each containing four layers. Each layer adds 32 feature channels and uses a bottleneck with 128 channels. Dropout rates are 0.2 within the dense layers and 0.5 in the classifier. This configuration contains approximately 0.9M parameters.

For the two foundation-model baselines, seegnificant and CBraMod, the encoders are initialized with official pretrained weights and paired with newly initialized classification heads. All encoder and classification-head weights are jointly fine-tuned on each downstream task using the shared training settings in Table~\ref{tab:Table.7}.

The seegnificant configuration uses an embedding dimension of 2, with one temporal and one spatial Transformer block. Its outputs are flattened and passed to a 128-unit hidden layer with ReLU activation and dropout with a rate of 0.5. A final linear layer produces one logit per target class. The complete implementation contains approximately 6.3M parameters under the reference setting.

CBraMod divides each recording channel into non-overlapping patches of 200 time samples and encodes them using 200-dimensional embeddings. The encoder has 12 layers, eight attention heads per layer, and a feed-forward hidden dimension of 800. For classification, the encoded patches are averaged over time to obtain 200 features per recording channel. These features are concatenated across channels and passed to a single linear layer with one output logit per target class. The complete implementation contains approximately 5.1M parameters under the reference setting.

\input{./tables/training_table}
The shared training and augmentation hyperparameters for all ten baselines and both NeuroSketch instantiations are summarized in Table~\ref{tab:Table.7}.

Across multiple datasets, we observed that decoding performance typically improves with an increased number of training epochs. Based on this observation, we set the number of training epochs to 500 in most experiments to fully optimize the model performance. However, for the \emph{Chisco} dataset, the accuracy reaches a plateau at around 80 epochs and shows no further improvement. Therefore, we limit training on \emph{Chisco} to 100 epochs.

The checkpoint with the highest validation accuracy is selected for test evaluation. Data splitting and augmentation procedures are described in Appendices~\ref{Data splitting} and~\ref{Data augmentation}, respectively.

\subsection{Compute Resources}\label{Compute resources}
We utilized an AMD EPYC 7663 56-core processor and eight NVIDIA A100 GPUs, each with 80 GB of memory. Section~\ref{roadmap} outlines 21 distinct models, each evaluated on 84 experiments that together span the entire set of subjects across eight decoding tasks. In total, this results in 1,764 experiments. Each experiment is executed on an A100 GPU and requires, on average, one hour of training time. Section~\ref{Experiments} evaluates 12 distinct models, each trained with three cross-validation folds, resulting in a total of 3,024 experiments. On average, each experiment requires approximately 50 minutes of training time. Together, the roadmap study and main evaluation comprise 4,788 training runs, or nearly 5,000 experiments.

\section{Additional Experimental Analyses}\label{Experimental Analysis}

\subsection{Scaling Behavior}\label{Scaling Behavior}
\begin{figure}[htbp]
   \centering
   \includegraphics[width=0.50\linewidth]{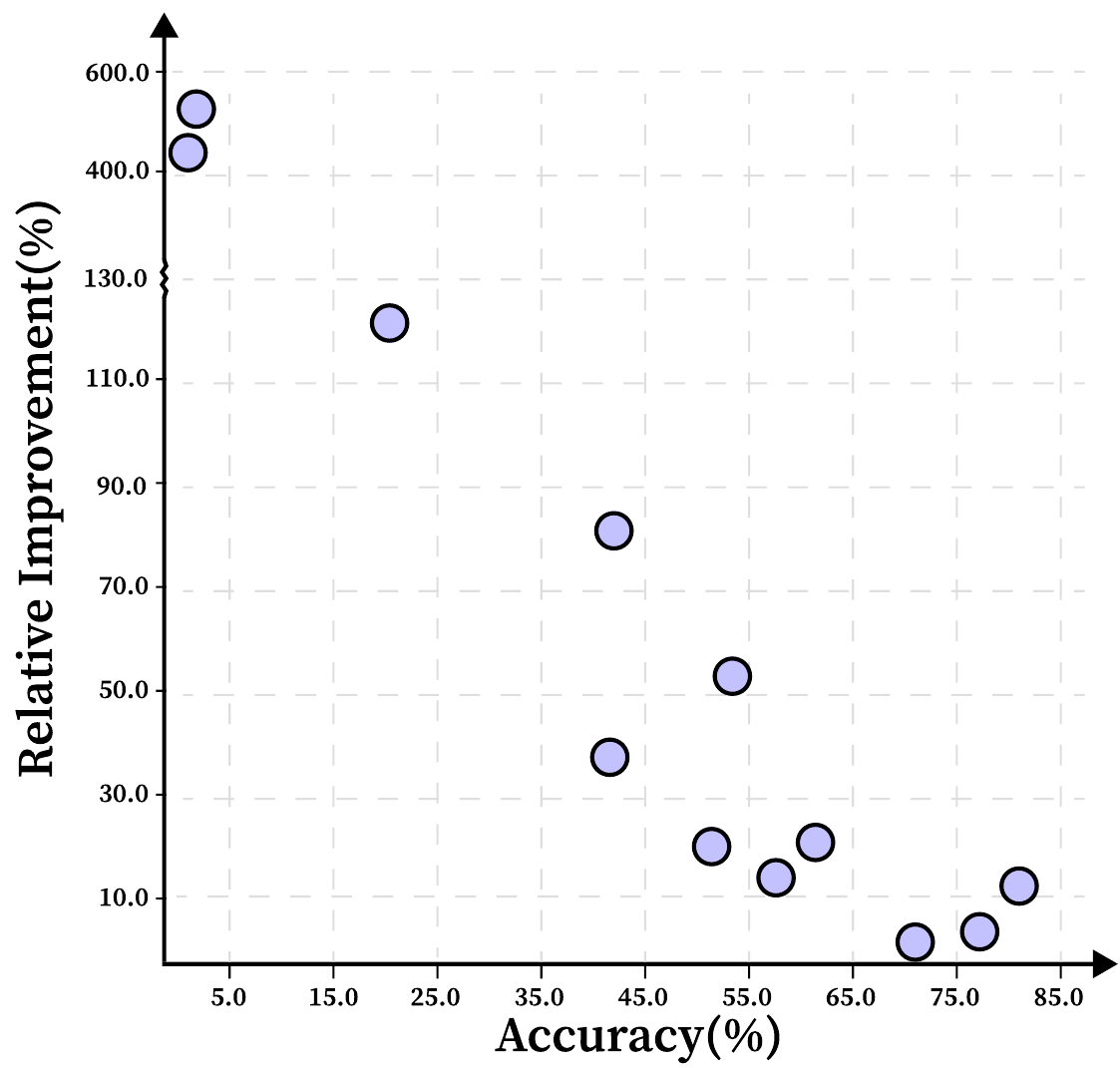}
   \caption{Scatter plot comparing the two NeuroSketch instantiations on the Du-IN dataset. The $x$ axis denotes the accuracy of NeuroSketch-Base for each subject and the $y$ axis represents the relative improvement achieved by NeuroSketch-Large.}
   \label{fig:Fig4}
\end{figure}

In this section, we analyze scaling behavior across the two NeuroSketch instantiations. As shown in Table~\ref{tab:Table.4}, NeuroSketch-Large achieves a clear performance improvement over NeuroSketch-Base on the Du-IN dataset, whereas the two models perform comparably on the other datasets. To further investigate this phenomenon, we examine subject-level performance on the Du-IN dataset, with the results summarized in Table~\ref{tab:Table.8}. We observe that NeuroSketch-Large outperforms NeuroSketch-Base across all subjects, although the extent of improvement varies. To examine how scaling gains vary across subjects, we plot relative accuracy improvement against NeuroSketch-Base accuracy for each subject (Figure~\ref{fig:Fig4}). Subjects with lower Base accuracy tend to show larger relative gains. Notably, for subjects with NeuroSketch-Base accuracy below 5\%, NeuroSketch-Large yields relative improvements of over 400\%. This indicates that increasing model capacity within the recipe effectively addresses more challenging tasks.

\input{./tables/scaling_behavior}

% As shown in Figure~\ref{fig:Fig4}, decoding accuracy and relative improvement are negatively correlated: subjects with lower baseline accuracy benefit more from scaling. Notably, for subjects with NeuroSketch-Base accuracy below 5\%, NeuroSketch-Large yields relative improvements of over 400\%, suggesting that the larger model is particularly effective in more challenging decoding scenarios. Moreover, all subjects show positive gains when moving from NeuroSketch-Base to NeuroSketch-Large, confirming that scaling consistently enhances decoding performance across the entire cohort.

\subsection{Representation Analysis of CNN-2D and Transformer‑Based Architectures}\label{Representation Analysis of CNN-2D and Transformer‑Based Architectures}
To further characterize the performance differences in Sections~\ref{Basic Architecture Study} and~\ref{Results Sec 3}, we compare the representations learned by NeuroSketch-Large and MedFormer.

Specifically, we examined the model representations from a rank‑based perspective, following the framework proposed by~\citet{yu2025understanding}. The rank of a feature representation reflects its expressive capacity: a higher rank indicates richer and more diverse features, while a lower rank implies excessive compression and loss of independent information. In a well‑organized hierarchy, the rank is expected to remain high or increase with depth, as deeper layers capture more abstract yet independent features. In our evaluation, since the CNN-2D-based model and the Transformer-based model achieved the overall best and worst performance, respectively, we conducted an analysis on NeuroSketch-Large (CNN-2D-based) and MedFormer (Transformer-based) to examine how the rank of the feature representations produced by each layer evolves across network depth.

For any layer of a given model, let the output representation of a sample be denoted as
$\mathbf{O}\in \mathbb{R}^{s\times d}$, where $s$ represents the sequence length and $d$ the feature dimension. We performed singular value decomposition (SVD) on $\mathbf{O}$ as $\mathbf{O}=\mathbf{U}\Sigma \mathbf{V}^{T}$,  where $\mathbf{U}$ and $\mathbf{V}$ are orthogonal matrices, and $\Sigma$ is a diagonal matrix whose diagonal entries $\sigma _1\ge \sigma _2\ge \cdots \ge \sigma _{\min \left\{ s,d \right\}}$ are the singular values. Although the algebraic rank—i.e., the number of nonzero singular values—serves as a strict measure of rank, real‑world data are often noisy. Therefore, we adopt a more practical numerical rank. For a tolerance $\varepsilon>0$, the $\varepsilon$-rank of $\mathbf{O}$ is defined as the number of singular values that are significant relative to the largest one:
\begin{equation}
   \varepsilon\text{-rank}\left( \mathbf{O} \right) =\left| \left\{ \left. i \right|\frac{\sigma _i\left( \mathbf{O} \right)}{\sigma _1\left( \mathbf{O} \right)}>\varepsilon \right\} \right|,
\end{equation}
Since the maximum rank (defined as $\min(s, d)$) varies across layers depending on $s$ and $d$, we further compute the $\varepsilon$-rank ratio, defined as the $\varepsilon$-rank relative to the maximum rank. This metric allows for a more intuitive comparison of how the effective rank evolves across layers. 
\begin{equation}
   \varepsilon\text{-rank ratio}=\frac{\varepsilon\text{-rank}\left( \mathbf{O} \right)}{\min \left( s, d \right)},
\end{equation}

For MedFormer, we extracted the $\varepsilon$-rank ratios of the embedding layer and each of its six encoder layers under three different patch lengths (5,10 and 20) to examine how patch length affects representational rank. For NeuroSketch‑Large, we analyzed the outputs of its three stem layers and all layers across the four feature‑extraction stages, transforming the height–width dimensions of each feature map into a single sequence length for a consistent rank analysis.

\begin{figure}[htbp]
   \centering
   \includegraphics[width=1.0\linewidth]{./figures/Fig6.pdf}
   \caption{Layer-wise $\varepsilon$-rank ratio analysis of MedFormer and NeuroSketch-Large.}
   \label{fig:Fig6}
\end{figure}

As shown in Figure~\ref{fig:Fig6}, from the two models' evolution of the $\varepsilon$-rank ratios, we can observe that (1) For multi-channel EEG signals, the embeddings are not as low‑rank as those typically observed in conventional time‑series~\citep{liang2025attention,yu2025understanding}. Specifically, for both MedFormer and NeuroSketch‑Large, when $\varepsilon \le 10^{-2.5}$, the $\varepsilon$‑rank of the embedding layer (for MedFormer) and the first stem layer (for NeuroSketch‑Large) remains above 60\% of the maximum possible rank, indicating that both models retain high‑dimensional representations rather than collapsing into a low‑rank subspace. 
Increasing MedFormer’s patch length from 5 to 20 also raises the maximum $\varepsilon$-rank ratio, suggesting that longer temporal patches yield higher-rank representations. These observations may be related to the information compression ratio at the initial embedding stage. For example, in MedFormer, the embedding stage applies a 2D convolution kernel of shape $[C, P]$, where $C$ is the number of input channels, and $P$ is the patch length. The convolution output is projected into an embedding space of dimension $D$. Consequently, the effective compression ratio is approximately $\frac{C \times P}{D}$. Given that brain signals typically involve a large number of channels (often 64 or more), this ratio is much higher than in univariate time‑series data. Increasing $P$ further amplifies the compression, producing embeddings with higher numerical rank which is consistent with our observations.

(2) For both NeuroSketch-Large and MedFormer, we observe distinct trends in how the $\varepsilon$-rank ratio evolves across layers. As the network depth increases, MedFormer tends to progressively reduce the rank of its feature representations, indicating excessive compression of the input information into a lower‑dimensional subspace. In contrast, NeuroSketch-Large exhibits the opposite trend: the rank of its representations gradually increases, approaching full rank in deeper layers. These contrasting trends imply that the two architectures differ fundamentally in how they organize their representation spaces. In this comparison, NeuroSketch-Large preserves higher-rank representations in deeper layers than MedFormer.
   
\subsection{Hyperparameter Analysis}\label{Hyperparameter Analysis}
% Since the architecture of nsk is based on cnn2d, we conduct hyperparameter analysis on the kernel size of conv and group number of group conv...
Since both NeuroSketch instantiations use the CNN-2D backbone, convolutional operations play a central role in their performance. To better understand their impact, we conducted a hyperparameter analysis focusing on two key factors: the kernel size and the number of groups in group convolutions.

\vpara{Kernel size.} The size of the convolutional kernel directly affects the receptive field and therefore controls the extent of local temporal–spatial dependencies that the model can capture. In NeuroSketch-Large, we adopt a kernel size of 3 as the default setting, since smaller kernels are generally more effective in capturing fine-grained dynamics while keeping the parameter overhead low. To further examine the effect of kernel size, we additionally evaluate larger kernels of 5 and 7 on the Du-IN dataset. The results are summarized in Table~\ref{tab:Table.9}.

\input{./tables/kernel_size}

These results indicate that larger kernel sizes lead to lower decoding accuracy, which aligns with our claim in Section~\ref{introduction} that neural decoding signals exhibit transient temporal dynamics. Larger kernels tend to capture longer-range temporal features, which may dilute short-term patterns that are critical for accurate decoding in this context. 

\vpara{Number of groups.}The number of groups in group convolution determines how the feature channels are partitioned and processed, which in turn affects both the computational cost and the representational capacity. Increasing the number of groups reduces the computational cost, since fewer channels are convolved together within each group. In NeuroSketch-Large, we set the default group number to 4. To further examine the effect of this parameter, we additionally evaluate group numbers of 2, 8, and 16 on the Du-IN dataset. The results are summarized in Table~\ref{tab:Table.10}.

\input{./tables/group_number}

The results indicate that setting the group number to 4 provides the most favorable configuration. Compared with using 2 groups, this setting achieves competitive decoding performance while reducing the computational cost. At the same time, it clearly outperforms larger group numbers of 8 and 16 in terms of decoding accuracy, highlighting that 4 groups strike an effective balance between efficiency and representational capacity.

\subsection{Component-wise Ablation Study}\label{Component-wise Ablation Study}
To systematically evaluate the independent contributions of key architectural components in the model instantiations, we conduct ablation experiments on the pooling strategy and residual connections using NeuroSketch-Base.

\vpara{Pooling strategy.} We compare three pooling methods: Generalized mean pooling (GeM) (used in the final model), average pooling, and max pooling. Table~\ref{tab:pooling_ablation} summarizes the results. 

\input{./tables/pooling_strategy.tex}
\input{./tables/residual_connections.tex}

GeM pooling yields consistent improvements over average pooling and substantially outperforms max pooling across all datasets. The learnable parameter in GeM allows adaptive aggregation of temporal and spatial features, which is beneficial for capturing discriminative patterns.
  
\vpara{Residual connections.}
Table~\ref{tab:residual_ablation} compares NeuroSketch-Base with and without residual connections. Removing residual connections reduces accuracy across the evaluated tasks, supporting their role in the final model.

\subsection{Cross-Subject Evaluation}\label{Cross-Subject Evaluation}
The cross-subject evaluation examines whether NeuroSketch retains its decoding performance when applied to participants unseen during training. NeuroSketch-Base is compared with EEG-Conformer and CBraMod on PhysioNet-MI and SEED-DV, using disjoint subjects for training and testing.

PhysioNet-MI is a four-class motor-imagery task derived from the publicly available EEG Motor Movement/Imagery Dataset~\citep{schalk2009eegmmidb}. The classes correspond to imagined movements of the left fist, right fist, both fists, and both feet. We use imagery runs 04, 06, 08, 10, 12, and 14. The 64-channel EEG recordings, originally sampled at 160 Hz, are resampled to 200 Hz and segmented into 4-second inputs containing 800 time samples per channel. Participants are assigned to disjoint training, validation, and test sets by sorting the subject directory names and selecting the first 70 participants for training, the next 19 for validation, and the next 20 for testing. This subject partition remains fixed across five training runs with random seeds 42, 43, 44, 45, and 46.

For SEED-DV, we evaluate 40-class visual concept classification using all 20 participants. Each input contains 62 EEG channels sampled at 200 Hz over 2 seconds, giving 400 time samples per channel. Four participants (20\%) are randomly held out for testing using a split seed of 42. The remaining 16 participants are partitioned into three groups for training and validation: each group serves as validation once, while the other two are used for training. All samples from a participant remain in the same partition, and the four test participants are fixed across the three folds. For each fold, the checkpoint with the highest validation accuracy is evaluated on the pooled samples of the four test participants. We report the mean and standard deviation of these three test accuracies.

\begin{table}[htbp]
\centering
\caption{\textbf{Cross-subject decoding accuracy} (mean $\pm$ standard deviation). Higher values indicate better performance.}
\label{tab:cross_subject}
\begin{tabular}{@{}lcc@{}}
\toprule
Model & PhysioNet-MI & SEED-DV \\
\midrule
EEG-Conformer & $0.537\pm0.004$ & $0.032\pm0.005$ \\
CBraMod & $0.613\pm0.010$ & $0.025\pm0.000$ \\
NeuroSketch-Base & $\mathbf{0.621}\pm0.008$ & $\mathbf{0.044}\pm0.011$ \\
\bottomrule
\end{tabular}
\end{table}

Table~\ref{tab:cross_subject} shows that NeuroSketch-Base achieves the highest mean accuracy among the three models on both tasks. On PhysioNet-MI, its accuracy reaches 62.1\%, compared with 61.3\% for CBraMod and 53.7\% for EEG-Conformer. On SEED-DV, the corresponding accuracies are 4.4\%, 2.5\%, and 3.2\%, respectively. The gains over the strongest baseline are therefore 0.8 and 1.2 percentage points on PhysioNet-MI and SEED-DV. These results extend the evaluation of NeuroSketch to subject shifts in both motor-imagery and visual decoding. Nevertheless, the low absolute accuracies on SEED-DV indicate that cross-subject visual decoding remains challenging in this setting.

\subsection{Robustness to Reduced Training Data}\label{Robustness to Reduced Training Data}
To examine sensitivity to the amount of labeled training data, we evaluate NeuroSketch-Base and SPaRCNet on OpenMIIR perception and imagery. Both tasks classify the twelve musical stimuli described in Appendix~\ref{Dataset Description}, using the inputs defined in Appendix~\ref{Data Preprocessing}. For each task, we retain 50\%, 75\%, or 100\% of the training split while keeping the validation and test sets unchanged. For each participant, the reduced training sets are randomly sampled using a fixed seed of 42. The 100\% setting provides the full-training-data reference for interpreting the reduced-data results. These percentages apply only to the training split, so a smaller fraction does not reduce the evaluation set. Nine participants are evaluated using the three-fold protocol in Appendix~\ref{Data splitting}. Accuracy is averaged across participants within each fold, and the reported mean and standard deviation are calculated across the three fold-level averages.

\begin{table}[htbp]
\centering
\caption{\textbf{OpenMIIR accuracy with reduced training data} (\%, mean $\pm$ standard deviation). The percentage in the first column applies only to the training split.}
\label{tab:low_data}
\setlength{\tabcolsep}{5pt}
\begin{tabular}{@{}lcccc@{}}
\toprule
 & \multicolumn{2}{c}{Perception} & \multicolumn{2}{c}{Imagery} \\
\cmidrule(lr){2-3}\cmidrule(lr){4-5}
Training data & SPaRCNet & NeuroSketch-Base & SPaRCNet & NeuroSketch-Base \\
\midrule
50\% & $86.4\pm5.2$ & $\mathbf{86.8}\pm5.2$ & $80.7\pm0.7$ & $\mathbf{83.2}\pm0.8$ \\
75\% & $94.8\pm2.9$ & $\mathbf{96.0}\pm2.4$ & $91.3\pm1.0$ & $\mathbf{93.2}\pm0.9$ \\
100\% & $96.9\pm0.7$ & $\mathbf{98.1}\pm1.0$ & $94.8\pm0.5$ & $\mathbf{97.0}\pm0.2$ \\
\bottomrule
\end{tabular}
\end{table}

Table~\ref{tab:low_data} shows that accuracy increases with the training fraction for both models on both tasks. NeuroSketch-Base achieves higher mean accuracy than SPaRCNet at every evaluated fraction. With 50\% of the training data, its accuracies reach 86.8\% for perception and 83.2\% for imagery. These exceed SPaRCNet by 0.4 and 2.5 percentage points, respectively. At 75\%, the corresponding gains are 1.2 and 1.9 percentage points. The larger gains on imagery are also present with the full training split, where NeuroSketch-Base leads by 2.2 percentage points. Thus, its advantage over SPaRCNet persists under reduced training data, although both models benefit substantially from additional labeled examples.

\subsection{Computational Efficiency}\label{Computational Efficiency}
To characterize the computational cost of the recipe's instantiation, we compare both NeuroSketch variants with EEGNet, EEG-Conformer, and CBraMod. All models are benchmarked on an NVIDIA A100 GPU using synthetic inputs with 64 recording channels and 2000 time points, a batch size of 64, and full numerical precision (FP32). Table~\ref{tab:efficiency} reports parameter counts together with training and inference throughput, measured in samples per second. Reporting both throughput measures distinguishes the cost of model fitting from the rate of prediction after training.

\begin{table}[htbp]
\centering
\caption{\textbf{Model size and computational throughput.} Throughput is measured in samples per second; higher values indicate faster processing.}
\label{tab:efficiency}
\begin{tabular}{@{}lrrr@{}}
\toprule
Model & Parameters (M) & Training & Inference \\
\midrule
EEGNet & 1.5 & 260 & 896 \\
EEG-Conformer & 0.2 & 95 & 1168 \\
CBraMod & 5.1 & 57 & 618 \\
NeuroSketch-Base & 1.4 & 130 & 1262 \\
NeuroSketch-Large & 4.2 & 103 & 1258 \\
\bottomrule
\end{tabular}
\end{table}

NeuroSketch-Base achieves the highest inference throughput among the five models, processing 1262 samples per second. Compared with CBraMod, it achieves 2.04$\times$ the inference throughput and 2.28$\times$ the training throughput. NeuroSketch-Large achieves a comparable inference throughput of 1258 samples per second, demonstrating high inference efficiency across both variants.

%% file: tables/Model_table.tex
\begingroup
\setlength{\tabcolsep}{1pt}   % 增加列间距
\renewcommand{\arraystretch}{1.0} % 调整行高
\setlength{\heavyrulewidth}{1pt}   % 顶部/底部线宽
\setlength{\lightrulewidth}{0.5pt} % 中间线宽

\begin{table}[htbp]
\centering
\small % 减小字体大小
\caption{\textbf{Model configuration for basic architecture study.}}
\label{tab:Table.5}
\NeuroSketchFitTable{\begin{tabular}{@{}cccc@{}} % 使用@{}消除边距，增加列宽
\toprule
\textbf{Model} & \textbf{\# Params (M)} & \textbf{\# Layers} & \textbf{\# Dimension} \\ 
\midrule
CNN-1D         & 30  & 20 & 1024 \\
CNN-2D         & 35  & 16 & 768  \\
GRU            & 25  & 4  & 512  \\
Transformer    & 20  & 8  & 512  \\
PatchTST       & 20  & 8  & 512  \\
iTransformer   & 20  & 8  & 512  \\
CNN-GRU        & 34  & 16 & 512  \\
CNN-Transformer & 32 & 20 & 768  \\
CNN-Transformer v2 & 36 & 20 & 1024 \\
\bottomrule
\end{tabular}}
\end{table}
\endgroup

%% file: tables/nsk_details.tex
\begingroup
\setlength{\tabcolsep}{1.8pt}   % Set the uniform column spacing
\renewcommand{\arraystretch}{1.1} % Reduce row height
\setlength{\heavyrulewidth}{1.0pt}   % Top and bottom line thickness
\setlength{\lightrulewidth}{0.5pt}    % Middle line thickness
\setlength{\arrayrulewidth}{0.5pt}    % Uniform thickness of horizontal lines

\begin{table}[htbp]
\centering % 让表格居中
\caption{\textbf{Detailed model configuration for NeuroSketch-Base and NeuroSketch-Large.}}
\label{tab:Table.6}
\NeuroSketchFitTable{\begin{tabular}{lcc}
\toprule
                                    & NeuroSketch-Base          & NeuroSketch-Large     \\ \midrule
\# Params (M)                            & 1.4                    & 4.2      \\
$D_\text{stage 1}$                  & 96                        & 96       \\
$D_\text{stage 2}$                  & 128                       & 144              \\
$D_\text{stage 3}$                  & 160                       & 256              \\
$D_\text{stage 4}$                  & 192                       & 384              \\
$d$                                 & 2                         & 3         \\
$G$                                 & 4                         & 4         \\
\bottomrule
\end{tabular}}
\end{table}

\endgroup

%% file: tables/training_table.tex
\begingroup
\setlength{\tabcolsep}{1.0pt}   % Set the uniform column spacing
\renewcommand{\arraystretch}{1.0} % Reduce row height
\setlength{\heavyrulewidth}{1.0pt}   % Top and bottom line thickness
\setlength{\lightrulewidth}{0.5pt}    % Middle line thickness
\setlength{\arrayrulewidth}{0.5pt}    % Uniform thickness of horizontal lines

\begin{table}[htbp]
\centering % 让表格居中
\caption{\textbf{Training hyperparameters.}}
\label{tab:Table.7}
\NeuroSketchFitTable{\begin{tabular}{lc}
\toprule
\multicolumn{1}{l}{Hyperparameters} & \multicolumn{1}{c}{Settings}    \\ \midrule
Epoch                               & 100(Chisco), 500(others)        \\
Batch size                          & 64                              \\
Seed                                & 42                              \\
Optimizer                           & AdamW                           \\
Learning rate                       & 1e-3 \\
Weight decay                        & 5e-2                            \\
Scheduler                           & Cosine                          \\
Warmup ratio                        & 0.1                             \\
Early stop ratio                    & 0.2                             \\
Random shift probability            & 0.5                             \\
Random shift ratio                  & 0.2                             \\
Add noise probability               & 0.1                             \\
Channel masking probability         & 0.5                             \\
Time masking probability            & 0.5                             \\
Mixup probability                   & 0.5                            \\
\bottomrule
\end{tabular}}
\end{table}

\endgroup

%% file: tables/scaling_behavior.tex
\begingroup
\setlength{\tabcolsep}{4pt}   % Set the uniform column spacing
\renewcommand{\arraystretch}{1.0} % Reduce row height
\setlength{\heavyrulewidth}{1.0pt}   % Top and bottom line thickness
\setlength{\lightrulewidth}{0.5pt}    % Middle line thickness
\setlength{\arrayrulewidth}{0.5pt}    % Uniform thickness of horizontal lines

\begin{table}[htbp]
\centering % 让表格居中
\caption{\textbf{Scaling comparison of NeuroSketch-Base and NeuroSketch-Large on the Du-IN dataset.} We report mean accuracy with standard deviation (computed over three cross-validation folds). RI denotes the relative improvement in accuracy of NeuroSketch-Large compared to NeuroSketch-Base.}
\label{tab:Table.8}
\NeuroSketchFitTable{\begin{tabular}{cccc}
\toprule
Subject & NeuroSketch-Base  & NeuroSketch-Large & RI\\ \midrule
1       & 0.626 {$\pm$0.018}   & 0.791 {$\pm$0.027}    &   26.37\%             \\
2       & 0.761 {$\pm$0.025}   & 0.798 {$\pm$0.007}    &   4.95\%             \\
3       & 0.081 {$\pm$0.025}   & 0.508 {$\pm$0.016}    &   529.34\%             \\
4       & 0.422 {$\pm$0.052}   & 0.765 {$\pm$0.033}    &   81.36\%             \\
5       & 0.774 {$\pm$0.012}   & 0.882 {$\pm$0.013}    &   13.90\%             \\
6       & 0.209 {$\pm$0.021}   & 0.465 {$\pm$0.016}    &   122.49\%             \\
7       & 0.409 {$\pm$0.019}   & 0.561 {$\pm$0.021}    &   37.28\%             \\
8       & 0.525 {$\pm$0.012}   & 0.630 {$\pm$0.021}    &   20.00\%             \\
9       & 0.540 {$\pm$0.030}   & 0.817 {$\pm$0.005}    &   51.36\%             \\
10      & 0.034 {$\pm$0.008}   & 0.183 {$\pm$0.003}    &   433.01\%             \\
11      & 0.721 {$\pm$0.013}   & 0.758 {$\pm$0.011}    &   5.09\%             \\
12      & 0.569 {$\pm$0.050}   & 0.668 {$\pm$0.010}    &   17.39\%             \\

\bottomrule
\end{tabular}}
\end{table}

\endgroup

%% file: tables/kernel_size.tex
\begingroup
\setlength{\tabcolsep}{4pt}   % Set the uniform column spacing
\renewcommand{\arraystretch}{1.1} % Reduce row height
\setlength{\heavyrulewidth}{1.0pt}   % Top and bottom line thickness
\setlength{\lightrulewidth}{0.5pt}    % Middle line thickness
\setlength{\arrayrulewidth}{0.5pt}    % Uniform thickness of horizontal lines

\begin{table}[htbp]
\centering % 让表格居中
\caption{\textbf{Results of different kernel sizes of NeuroSketch on the Du-IN dataset.} Results are reported as mean and standard deviation, computed across three distinct cross-validation folds.}
\label{tab:Table.9}
\NeuroSketchFitTable{\begin{tabular}{cccc}
\toprule
Kernel Size & Accuracy        & Precision       & F1 Score      \\ \midrule
3           & 0.651 $\pm$ 0.005    & 0.667 $\pm$0.003    & 0.647  $\pm$0.003  \\
5           & 0.557 $\pm$ 0.005 & 0.577 $\pm$ 0.006 & 0.550 $\pm$ 0.005 \\
7           & 0.427 $\pm$ 0.002 & 0.456 $\pm$ 0.002 & 0.420 $\pm$ 0.003 \\
\bottomrule
\end{tabular}}
\end{table}

\endgroup

%% file: tables/group_number.tex
\begingroup
\setlength{\tabcolsep}{4pt}   % Set the uniform column spacing
\renewcommand{\arraystretch}{1.1} % Reduce row height
\setlength{\heavyrulewidth}{1.0pt}   % Top and bottom line thickness
\setlength{\lightrulewidth}{0.5pt}    % Middle line thickness
\setlength{\arrayrulewidth}{0.5pt}    % Uniform thickness of horizontal lines

\begin{table}[htbp]
\centering % 让表格居中
\caption{\textbf{Effect of the number of convolution groups on NeuroSketch performance on Du-IN.} Results are reported as mean and standard deviation, computed across three distinct cross-validation folds.}
\label{tab:Table.10}
\NeuroSketchFitTable{\begin{tabular}{cccc}
\toprule
\#Group & Accuracy & Precision & F1 Score \\ \midrule
2       & 0.651 $\pm$ 0.005    & 0.667 $\pm$0.003    & 0.647  $\pm$0.003  \\ 
4       & 0.652 $\pm$ 0.002    & 0.666 $\pm$0.003    & 0.648  $\pm$0.002  \\
8       & 0.642 $\pm$ 0.002    & 0.658 $\pm$0.003    & 0.638  $\pm$0.002  \\
16      & 0.642 $\pm$ 0.001    & 0.656 $\pm$0.003    & 0.637  $\pm$0.002  \\

\bottomrule
\end{tabular}}
\end{table}
\endgroup

%% file: tables/pooling_strategy.tex
\begingroup
\setlength{\tabcolsep}{4pt}
  \renewcommand{\arraystretch}{1.1}
  \setlength{\heavyrulewidth}{1.0pt}
  \setlength{\lightrulewidth}{0.5pt}
  \setlength{\arrayrulewidth}{0.5pt}

  \begin{table}[htbp]
  \centering
  \caption{\textbf{Ablation study on pooling strategies.} We compare GeM pooling (used in NeuroSketch) with average and max pooling using NeuroSketch-Base. Results are reported as mean and standard deviation across three runs.}
  \label{tab:pooling_ablation}
  \NeuroSketchFitTable{\begin{tabular}{lccccc}
  \toprule
  Pooling & Chisco-R & Chisco-I & OpenMIIR-P & OpenMIIR-I & ThingsEEG \\
  \midrule
  GeM (Ours) & \textbf{0.111 $\pm$ 0.001} & \textbf{0.103 $\pm$ 0.002} & \textbf{0.981 $\pm$ 0.010} & \textbf{0.970 $\pm$ 0.002} & \textbf{0.196 $\pm$ 0.002} \\
  Average    & 0.103 $\pm$ 0.002 & 0.095 $\pm$ 0.003 & 0.977 $\pm$ 0.013 & 0.969 $\pm$ 0.002 & 0.192 $\pm$ 0.001 \\
  Max        & 0.085 $\pm$ 0.002 & 0.077 $\pm$ 0.006 & 0.972 $\pm$ 0.007 & 0.946 $\pm$ 0.001 & 0.170 $\pm$ 0.003 \\
  \bottomrule
  \end{tabular}}
  \end{table}
\endgroup

%% file: tables/residual_connections.tex
\begingroup
\setlength{\tabcolsep}{4pt}
  \renewcommand{\arraystretch}{1.1}
  \setlength{\heavyrulewidth}{1.0pt}
  \setlength{\lightrulewidth}{0.5pt}
  \setlength{\arrayrulewidth}{0.5pt}

  \begin{table}[htbp]
  \centering
  \caption{\textbf{Ablation study on residual connections.} We evaluate NeuroSketch-Base with and without residual connections. Results are reported as mean and standard deviation across three runs.}
  \label{tab:residual_ablation}
  \NeuroSketchFitTable{\begin{tabular}{lccccc}
  \toprule
  Residual & Chisco-R & Chisco-I & OpenMIIR-P & OpenMIIR-I & ThingsEEG \\
  \midrule
  w/ (Ours) & \textbf{0.111 $\pm$ 0.001} & \textbf{0.103 $\pm$ 0.002} & \textbf{0.981 $\pm$ 0.010} & \textbf{0.970 $\pm$ 0.002} & \textbf{0.196 $\pm$ 0.002} \\
  w/o       & 0.084 $\pm$ 0.002 & 0.073 $\pm$ 0.006 & 0.910 $\pm$ 0.009 & 0.923 $\pm$ 0.004 & 0.110 $\pm$ 0.001 \\
  \bottomrule
  \end{tabular}}
  \end{table}
\endgroup